\documentclass{article}

\usepackage{abstract} 
\usepackage[top=1in,bottom=1in,left=1in,right=1in]{geometry}
\usepackage[bf,small,pagestyles]{titlesec}
\usepackage{graphicx} 
\usepackage{float}
\usepackage{multirow}
\usepackage{booktabs}
\usepackage{caption}
\usepackage{indentfirst}
\usepackage{subcaption}
\usepackage{amssymb}
\usepackage{amsmath}
\usepackage{amsthm}
\usepackage{cases}
\usepackage{xcolor}
\usepackage{algorithm} 
\usepackage{algpseudocode} 
\usepackage{caption}
\usepackage{hyperref}
\usepackage{comment}
\usepackage{csquotes}
\usepackage[figure,table]{hypcap}
\hypersetup{
	bookmarksnumbered,
	pdfstartview={FitH},
	citecolor={black},
	linkcolor={black},
	urlcolor={black},
	pdfpagemode={UseOutlines}
}
\theoremstyle{theoremStyle}
\theoremstyle{assumptionStyle}
\newtheorem{assumption}{Assumption} 
\newtheorem{theorem}{Theorem}
\newtheorem{definition}{Definition}
\newtheorem{lemma}{Lemma}
\newtheorem{proposition}{Proposition}
\newtheorem{remark}{Remark}
\usepackage{multicol}

\usepackage{color}
\usepackage{ulem} 
\usepackage{authblk}

\begin{document}
\title{\textbf{XNet-Enhanced Deep BSDE Method and Numerical Analysis}}
\author[1]{Xiaotao Zheng\thanks{Email: \texttt{20234013002@stu.suda.edu.cn}}}
\author[1]{Xingye Yue\thanks{Email: \texttt{xyyue@suda.edu.cn}}}
\author[2,3]{Zhihong Xia\thanks{Co-Corresponding author: \texttt{xia@math.northwestern.edu}}}
\author[2]{Xin Li\thanks{Co-corresponding author: \texttt{xinli2023@u.northwestern.edu}}}

\affil[1]{Center for Financial Engineering, Soochow University, Suzhou 215008, Jiangsu, China}
\affil[2]{Institute of Advanced Research, Great Bay University, Dongguan 523808, Guangdong, China}
\affil[3]{Department of Mathematics, Northwestern University, Evanston 60208, IL, USA}

\date{}

\phantomsection
\addcontentsline{toc}{section}{Title}
\maketitle
	
\phantomsection
\addcontentsline{toc}{section}{Abstract}

\begin{abstract}
%Solving high-dimensional semilinear parabolic partial differential equations (PDEs) challenges traditional numerical methods due to the "curse of dimensionality." Deep learning, particularly through the Deep BSDE method, offers a promising alternative by leveraging neural networks' capability to approximate high-dimensional functions. 
%However, the theoretical foundation of Deep BSDE method remains incomplete for many practically important PDEs that violate standard Lipschitz conditions, such as Allen-Cahn equations with cubic nonlinearity and Hamilton-Jacobi-Bellman (HJB) equations with quadratic gradient growth.
%This paper makes two key contributions. First, we extend the convergence theory of the Deep BSDE method to two important classes of non-Lipschitz PDEs: Allen-Cahn type equations and HJB type equations. Through novel truncation techniques and careful analysis of bounded solution properties, we provide the first rigorous theoretical justification for the empirical success of Deep BSDE on these equation classes. Second, we introduce a novel network architecture, XNet, which significantly enhances the computational efficiency and accuracy of the Deep BSDE method. XNet demonstrates superior approximation capabilities with fewer parameters, addressing the trade-off between approximation and optimization errors found in existing methods. We detail the implementation of XNet within the Deep BSDE framework and present results that show marked improvements in solving high-dimensional PDEs, potentially setting a new standard for such computations.	

Semilinear parabolic partial differential equations (PDEs) are fundamental to modeling complex dynamical systems across scientific domains. The Deep Backward Stochastic Differential Equation (BSDE) method is a promising approach for high-dimensional PDEs; however, existing convergence results apply only to globally Lipschitz generators, excluding important cases such as Allen--Cahn and Hamilton--Jacobi--Bellman (HJB) equations.

This paper presents both a theoretical and a computational advance for Deep BSDE methods. Theoretically, we establish the convergence theory for non--Lipschitz generators--covering Allen--Cahn equations with cubic nonlinearity and HJB equations with quadratic gradient growth--based on a bounded double--well lemma and a truncated-BSDE analysis within the Bouchard--Touzi--Zhang theory. Computationally, we instantiate the framework with XNet, a shallow architecture with $\mathcal O(L)$ parameters that preserves strong approximation while substantially reducing optimization and computational cost.
Numerical experiments on 100--dimensional PDEs corroborate the predicted convergence behavior and demonstrate significant efficiency gains over standard feedforward implementations.

%This paper establishes convergence theory for Deep BSDE methods applied to non-Lipschitz generators and introduces XNet, a parameter-efficient neural architecture for high-dimensional PDE computation. We prove convergence for two important PDE classes previously excluded from theoretical analysis: Allen-Cahn equations with generator $f(t,x,y,z) = y - y^3$ and Hamilton-Jacobi-Bellman equations with $f(t,x,y,z) = -\frac{1}{2}|z|^2$. For Allen-Cahn equations, we exploit boundedness properties of double-well dynamics to establish local Lipschitz conditions. our main contribution is proving that the Deep BSDE scheme converges to the truncated BSDE system, thereby establishing convergence to the BSDE system through existing truncation error bounds. We introduce XNet, characterized by shallow architecture with $\mathcal{O}(L)$ parameter complexity compared to $\mathcal{O}(HL^2)$ for standard feedforward networks. This addresses the approximation-optimization trade-off critical in multi-network Deep BSDE implementations. We detail the implementation of XNet within the Deep BSDE framework and present results that show marked improvements in solving high-dimensional PDEs, potentially setting a new standard for such computations.

\end{abstract}

\section{Introduction}
Semilinear parabolic partial differential equations (PDEs) play a crucial role in modeling complex dynamic systems across various scientific domains, from financial mathematics to biological processes. Consider the general form:

\begin{equation}\label{bxxpwxpde}
\left\{
\begin{aligned}
& \partial_t u(t,x)
+ \tfrac{1}{2}\operatorname{Tr}\!\big(\sigma \sigma^{\!T}(t,x)\, D_x^2 u(t,x)\big)
+ \mu(t,x)\!\cdot\!\nabla_x u(t,x) \\
&\qquad + f\!\left(t,x, u(t,x), \sigma^{\!T}(t,x)\nabla_x u(t,x)\right) = 0,\quad (t,x)\in[0,T)\times\mathbb{R}^d,\\
& u(T,x)=g(x),\quad x\in\mathbb{R}^d.
\end{aligned}
\right.
\end{equation}

Although traditional numerical methods such as the Finite Difference Method (FDM) and Finite Element Method (FEM) perform well in handling low-dimensional cases ($d \leq 3$), they struggle to solve high-dimensional problems due to the "curse of dimensionality" {\cite{weishu1}}.

Recent deep--learning approaches--PINNs~\cite{PINN_2024_1,Mishra,PINN,Shin,PINN_2024_2}, 
Deep Galerkin~\cite{DGM_2024,DGM}, and Deep Ritz~\cite{Ritz_2024,Ritz}--have shown promise for high--dimensional PDEs.
However, sampling strategies become challenging in very high dimensions (e.g., $d\!\ge\!100$).
This motivates alternative deep learning approaches based on stochastic processes (Monte Carlo sampling), which include the Deep Backward Stochastic Differential Equation (BSDE) method {\cite{E2017}}, Deep Splitting {\cite{Beck2}}, Deep Backward Dynamic Programming (DBDP) method {\cite{DBDP}} and recent martingale-based approaches \cite{ cai2024martingale,cai2024soc}. 
Among these, the Deep BSDE method is a widely adopted technique for handling such highly high--dimensional PDEs (\ref{bxxpwxpde}).
This method leverages neural networks to approximate the gradient of the solutions to high--dimensional semilinear parabolic PDEs, utilizing the capacity of these networks to manage the stochastic nature of the solutions.

Despite its innovative approach and empirical success, the Deep BSDE method faces fundamental theoretical and computational limitations. Convergence analysis requires the generator function $f$ to satisfy global Lipschitz conditions \cite{Zhang2008time, Convergence-of-the-deep-BSDE-method}.  This excludes important PDE classes arising in applications, notably Allen-Cahn equations with cubic nonlinearity ($f = y - y^3$) and Hamilton-Jacobi-Bellman (HJB) equations with quadratic gradient growth ($f = -\frac{1}{2}|z|^2$). While numerical convergence has been observed for such equations \cite{ han2018solving,E2017}, theoretical justification remained absent.
From a practical perspective, the method requires careful balance of approximation, generalization, and optimization errors through appropriate network architecture design. Standard feedforward networks, despite universal approximation properties \cite{weishu}, face scalability limitations in the multi-network Deep BSDE framework.

In this context, we introduce two major contributions to address these theoretical and practical challenges. First, we provide the rigorous convergence analysis for Deep BSDE method applied to non-Lipschitz PDEs, specifically Allen-Cahn type equations and HJB type equations. 
For Allen-Cahn equations, we exploit boundedness properties of double-well potential dynamics to establish local Lipschitz conditions. For HJB equations, we prove convergence through systematic analysis within the Bouchard-Touzi-Zhang framework \cite{zhang2017backward}, demonstrating that the Deep BSDE scheme converges to truncated BSDE systems. This theoretical advancement expands the class of PDEs for which Deep BSDE methods have guaranteed convergence.
Second, we introduce XNet \cite{XNet1, XNet2}, a parameter-efficient neural architecture designed based on Cauchy's approximation theorem. XNet achieves $\mathcal{O}(L)$ parameter complexity compared to $\mathcal{O}(HL^2)$ for traditional feedforward networks, dramatically reducing computational burden while maintaining superior approximation capabilities. This addresses the approximation-optimization trade-off critical in multi-network Deep BSDE implementations.
We provide comprehensive numerical validation through 100-dimensional Allen-Cahn equations and nonlinear financial derivative pricing problems. The XNet-enhanced Deep BSDE method demonstrates superior computational efficiency and accuracy, with clearer convergence behavior observable across both test cases. Notably, for Allen-Cahn equations in continuous-time implementation, XNet enables observation of convergence rates approaching 1.6, which remain obscured when using standard feedforward architectures.

\paragraph{Main contributions.}
This work delivers both a theoretical and a computational advance for Deep BSDE. 
{Theoretically}, we develop a convergence framework for non-Lipschitz generators (Allen--Cahn with cubic nonlinearity; HJB with quadratic gradient growth). 
For Allen--Cahn, we exploit a bounded double-well lemma to recover local Lipschitz control; for HJB, we combine truncation with the Bouchard--Touzi--Zhang (BTZ) reformulation and {identify a martingale target} that the learner must approximate. 
The resulting error bounds are {architecture-agnostic}: they depend only on the accuracy of the target approximation at each time layer (time-averaged $Z$ under Lipschitz generators vs.\ a martingale target for quadratic HJB), not on any particular network. 
{Computationally}, we {instantiate} the framework with XNet~\cite{XNet1,XNet2}, a compact architecture motivated by Cauchy-type approximation that achieves strong accuracy with substantially fewer parameters, thereby reducing the optimization burden inherent in the multi-network Deep BSDE setting. 
Experiments on 100-dimensional benchmarks corroborate the predicted convergence behavior and demonstrate significant efficiency gains.

\begin{enumerate}
\item \textbf{Convergence analysis beyond Lipschitz.}
A rigorous framework covering Allen--Cahn (cubic nonlinearity) and HJB (quadratic gradient growth), built on a bounded double-well lemma and a truncated--BTZ analysis.

\item \textbf{Efficient instantiation and verification.}
XNet serves as an efficient instantiation that reduces the {target-approximation error} under fixed compute; 100D Allen--Cahn and nonlinear pricing tests show lower errors, faster runtimes, and clearer convergence behavior than standard feedforward implementations.
\end{enumerate}

\textbf{Outline of the article}
The remainder is organized as follows. Section \ref{s2} reviews the Deep BSDE method. Section \ref{s3} establishes convergence theory for Deep BSDE methods, extending the analysis to non-Lipschitz generators including Allen-Cahn and HJB equations. Sections \ref{s4} and \ref{s5} present discrete-time and continuous-time implementations, respectively, demonstrating XNet's superior approximation capabilities and convergence properties. Section 6 concludes the paper.

\section{Deep BSDE Method}\label{s2}
In this section, we begin by introducing the BSDE system related to semilinear parabolic partial differential equations (PDEs). Subsequently, we will present the Deep BSDE (DBSDE) method proposed by E et al. {\cite{E2017}}.

\subsection{Forward Backward Stochastic Differential Equation (FBSDE)}
Following the seminal work of Peng \cite{Peng1,Peng2}, there exists a well-established connection between the semilinear parabolic PDE \eqref{bxxpwxpde} and backward stochastic differential equations (BSDEs). This probabilistic representation forms the theoretical foundation for the Deep BSDE method.

Consider a filtered probability space $(\Omega,\mathcal{F},\mathbb{P})$ equipped with a $d$-dimensional standard Brownian motion $W=\left(W^{(1)}, \ldots, W^{(d)}\right):[0, T] \times \Omega \rightarrow \mathbb{R}^d$, where $\{\mathcal{F}_t\}_{t \in [0,T]}$ denotes the natural filtration generated by $W$. Let $X=\left(X^{(1)}, \ldots, X^{(d)}\right):[0, T] \times \Omega \rightarrow \mathbb{R}^d$, $Y:[0, T] \times \Omega \rightarrow \mathbb{R}$, and $Z:[0, T] \times \Omega \rightarrow \mathbb{R}^d$ be $\mathcal{F}$-adapted stochastic processes satisfying the following forward-backward stochastic differential equation (FBSDE) system:
\begin{numcases}{}\label{BSDE}
	X_t=\xi+\int_0^t \mu\left(s, X_s\right) \mathrm{d} s+\int_0^t \sigma\left(s, X_s\right) \mathrm{d} W_s,  \label{Xt}\\
	Y_t=g\left(X_T\right)+\int_t^T f\left(s, X_s, Y_s, Z_s\right) \mathrm{d} s-\int_t^T\left(Z_s\right)^T \mathrm{~d} W_s. \label{Yt}
\end{numcases}
Under certain conditions, this FBSDE system admits a unique solution and provides the probabilistic representation of the PDE solution. Specifically, for any $t \in[0, T]$, the following equation holds almost surely in probability,
\begin{equation}\label{YZ}
	\left\{ {\begin{aligned}
			& {{Y_t} = u\left( {t,{X_t}} \right)},\\
			& {{Z_t} = {\sigma ^ {T} }\left( {t,{X_t}} \right)\nabla u\left( {t,{X_t}} \right)}.
	\end{aligned}} \right.
\end{equation}
This connection enables the reformulation of the deterministic PDE-solving problem into a stochastic framework. Specifically, if the $d$-dimensional process $\{X_t\}_{t \in[0, T]}$ satisfies the forward SDE~\eqref{Xt}, then the PDE solution satisfies the following integral representation:
\begin{equation}\label{BSDEeq}
	\begin{aligned}
		u(t,{X_t}) - u(0,\xi)  =  &- \int_0^t {f\left( {s,X_s,u(s,X_s),{\sigma ^{{T}}}(s,X_s)\nabla u(s,X_s)} \right)} ds \\
		& +\int_0^t {{{[\nabla u(s,X_s)]}^T}\sigma (s,X_s)d{W_s}} .
	\end{aligned}
\end{equation}

\subsection{Implementation of the Deep BSDE Method}
After adapting the solution of PDE (\ref{bxxpwxpde}) to the SDE (\ref{BSDEeq}) by BSDE theory, given a partitioning of the interval $[0,T]$, $\pi:0=t_0<t_1<\dots<t_N=T$ with $\Delta t_n=t_{n+1}-t_n$, the solution at each time step can be approximated with the Euler-Maruyama scheme,
%\begin{scriptsize}
\begin{equation}\label{ls-BSDEeq}
	\begin{aligned}
		u\left( {{t_{n + 1}},{X_{{t_{n + 1}}}}} \right) - u\left( {{t_n},{X_{{t_n}}}} \right)   \approx & - f\left( {{t_n},{X_{{t_n}}},u\left( {{t_n},{X_{{t_n}}}} \right),{\sigma ^{{T}}}\left( {{t_n},{X_{{t_n}}}} \right)\nabla u\left( {{t_n},{X_{{t_n}}}} \right)} \right)\Delta {t_n}\\
		& + {{\left[ {\nabla u\left( {{t_n},{X_{{t_n}}}} \right)} \right]}^{\rm{T}}}\sigma \left( {{t_n},{X_{{t_n}}}} \right)\Delta {W_n},n = 0,1,...,N - 1,
	\end{aligned}
\end{equation}
where $\Delta W_n=W_{t_{n+1}}-W_{t_n}$ represents the Brownian motion increment over the time interval $[t_n, t_{n+1}]$.
%\end{scriptsize}
To achieve a globally approximate scheme, neural networks can be incorporated into the forward discretization process (\ref{ls-BSDEeq}). The first step towards this is to obtain training data by sampling $M$ independent paths ${\left\{ {X_{{t_n}}^m} \right\}_{0 \le n \le N}^{m = 1,2, \ldots, M}}$, where $\left\{X_{t_0}^m\right\}^{m=1, \ldots, M}=\xi$. The critical step next is employing the neural network parameters ${\theta _{{u_0}}}$, ${\theta _{\nabla {u_0}}}$ to approximate the solution and the gradient function at $t=t_0$ respectively. Let $\theta_n$ represent all network parameters approximating the gradient function $\nabla u(t,x)$ by the neural network at time $t=t_n$, where $n = 1,2,..., N - 1$.  
With the above approximations, the total set of parameters is $\theta=\left\{{\theta _{{u_0}}}, {\theta _{\nabla {u_0}}}, \theta_1, \theta_2, \ldots, \theta_{N-1}\right\}$. The equation (\ref{ls-BSDEeq}) can be rewritten as follows.
%\begin{scriptsize}
\begin{equation}\label{nn-ls-BSDEeq}
	\begin{aligned}
		\hat u\left( {{t_{n + 1}},X_{{t_{n + 1}}}^m} \right) & - \hat u\left( {{t_n},X_{{t_n}}^m} \right)  =  - f\left( {{t_n},X_{{t_n}}^m,\hat u\left( {{t_n},X_{{t_n}}^m} \right),{\sigma ^{{T}}}\left( {{t_n},X_{{t_n}}^m} \right)\nabla u_{{t_n}}^\theta \left( {{t_n},X_{{t_n}}^m} \right)} \right)\Delta {t_n}\\
		& + {\left[ {\nabla u_{{t_n}}^\theta \left( {{t_n},X_{{t_n}}^m} \right)} \right]^{\rm{T}}}\sigma \left( {{t_n},X_{{t_n}}^m} \right)\Delta W_{{t_n}}^m,m = 1,2,...,M,n = 0,1,...,N - 1.
	\end{aligned}
\end{equation}
%\end{scriptsize}
In particular, when $t=t_0$, we have
\begin{equation}\label{t0-nn-ls-BSDEeq}
	\begin{aligned}
		\hat u\left( {{t_1},X_{{t_1}}^m} \right) = {\theta _{{u_0}}} - f\left( {{t_0},\xi ,{\theta _{{u_0}}},{\sigma ^T}\left( {{t_0},\xi } \right){\theta _{\nabla {u_0}}}} \right)\Delta {t_0} + {\left[ {{\theta _{\nabla {u_0}}}} \right]^{\rm{T}}}\sigma \left( {{t_0},\xi } \right)\Delta W_{{t_0}}^m,m = 1,2,...,M.
	\end{aligned}
\end{equation}
By applying the globally approximate scheme (\ref{nn-ls-BSDEeq}), an approximate value of $u\left(t_N, X_{t_N}^m\right)$, denoted as $\hat{u}\left(t_N, X_{t_N}^m\right)$, can be output, where $m=1,2, \ldots, M$. The matching of a given terminal condition defines the expected loss function,
\begin{equation}\label{BSDE-LOSS}
	\begin{aligned}
		l(\theta ) = \frac{1}{M}\sum\limits_{m = 1}^M {{{\left| {g\left( {{X_T^m}} \right) - \hat u\left( {{t_N},X_{{t_N}}^m} \right)} \right|}^2}} .
	\end{aligned}
\end{equation}
%Network parameters are optimized using algorithms such as SGD \textsuperscript{\cite{SGD}}, L-BFGS-B \textsuperscript{\cite{LBFGSB1,LBFGSB2}}, Adagrad \textsuperscript{\cite{ADAGRAD}}, and Adam \textsuperscript{\cite{ADAM}} algorithms. 
Through optimizing the network parameters, it becomes evident that an approximate solution
${\theta _{{u_0}}}$ for $u(0, \xi)$ can be obtained. 
%In the work of Weinan E et al., they opted to use the Adam algorithm.

\begin{figure*}[!hbtp]
	\begin{center}
		\includegraphics[angle=0,width=6in]{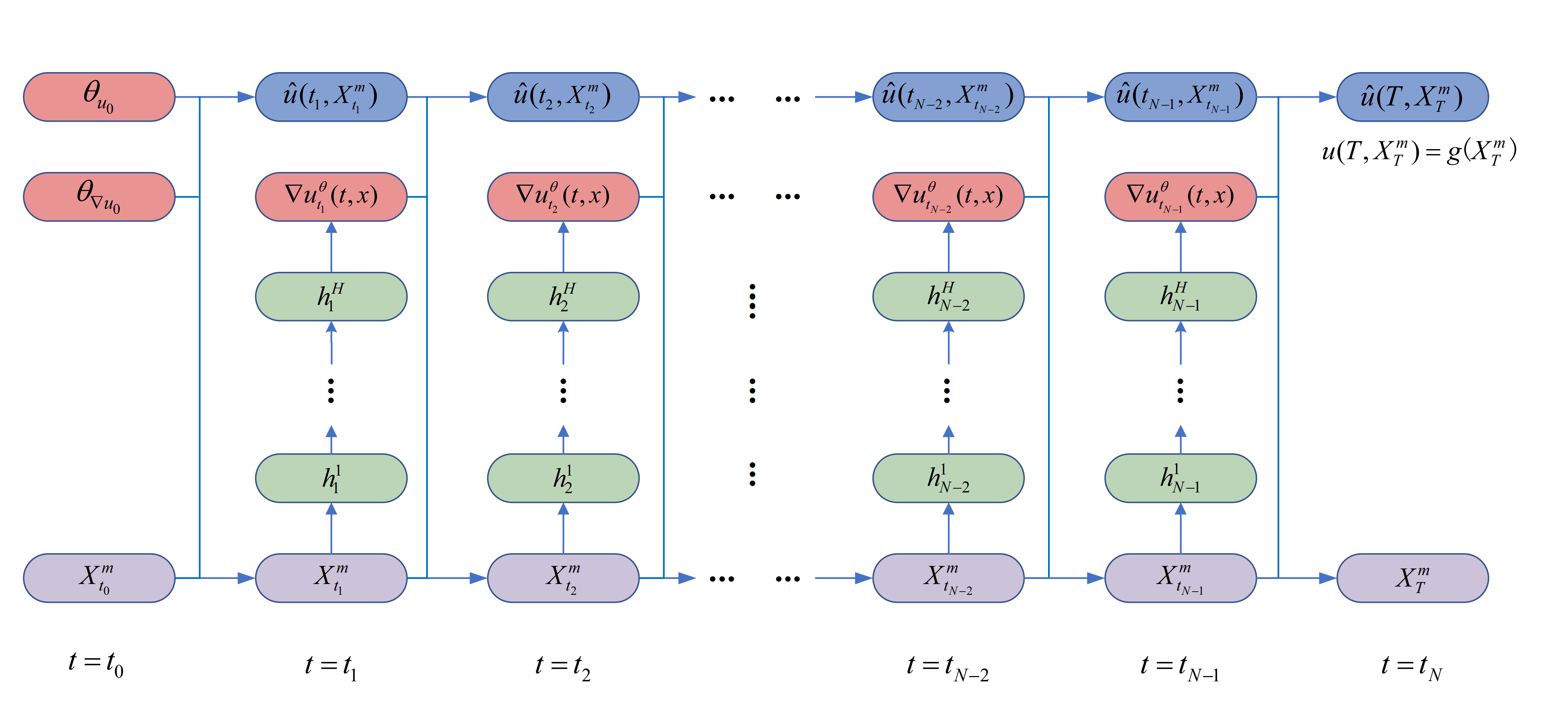}		
	\end{center}
	\vspace{-0.7cm}
	\caption{The neural network architecture for Deep BSDE method. The network consists of multiple ($N-1$) sub-networks, with each sub-network corresponding to a time interval. Each sub-network has $H$ hidden layers. It should be noted that in addition to these, ${\theta _{{u_0}}}$ and ${\theta _{\nabla {u_0}}}$ are also network parameters that need to be optimized.}
	\label{fig0}
\end{figure*}

The Deep BSDE method involves three error types: approximation, generalization, and optimization errors. Proper network architecture design minimizes approximation errors and optimization cost. Figure \ref{fig0} illustrates the multi-network architecture, highlighting parameter optimization complexity in discrete-time settings.

\subsection{Network Architecture for the Deep BSDE Method: Feedforward Neural Networks and XNet}

The original Deep BSDE method \cite{E2017} employed feedforward neural networks (FNNs) for gradient approximation. As demonstrated in Section \ref{s3}, the convergence properties of the Deep BSDE method critically depend on the approximation capabilities of the underlying neural network architecture. Traditional FNNs struggle to balance approximation capability with optimization cost. We propose replacing FNNs with shallow XNet architectures, demonstrating that XNet provides enhanced approximation efficiency while maintaining computational tractability.

The mathematical framework of feedforward neural networks (FNNs) is:
\begin{equation}\label{FNN}
	\begin{aligned}
    Y_{\text{FNN}}^\theta(X;W,b) = \sigma_H \!\left(
{W}_{H} 
\sigma_{H-1} \!\left(
\cdots
\sigma_1\!\left(
{W}_{1} X + {b}_{1}
\right)
+ \cdots + {b}_{H-1}
\right)
+ {b}_{H}
\right),
	\end{aligned}
\end{equation}
with parameter sets
$W = \{W_1, W_2, \dots, W_H \}, \quad b = \{b_1, b_2, \dots, b_H \}.$
Here, $X \in \mathbb{R}^d$ is the input, $W_i \in \mathbb{R}^{n_i \times n_{i-1}}$ and $b_i \in \mathbb{R}^{n_i}$ denote the weight matrices and bias vectors for the $i$-th layer, respectively, with $n_i$ representing the number of neurons in layer $i$. The activation functions $\sigma_i$ (commonly sigmoid, tanh, or ReLU) introduce the requisite nonlinearity for function approximation.

The feedforward networks employed in the original Deep BSDE implementation typically consist of $H$ hidden layers with $L$ neurons each, yielding a total parameter complexity of $\mathcal{O}(HL^2)$. While the universal approximation theorem \cite{weishu} establishes that sufficiently large FNNs guarantee approximation capability, the parameter growth presents significant computational challenges, particularly in the multi-network Deep BSDE framework where optimization complexity scales dramatically with network size.

\begin{theorem}[Cauchy Approximation Theorem; Theorem~1 in Xia~\cite{XNet1}]
\label{Cauchy_approximation_theorem}
	
Let $f(z^1,\dots,z^d)$ be analytic on an open set $U\subset\mathbb C^d$,
and let $M\subset U$ be compact.
Then for any $\varepsilon>0$, there exist points
$(\xi_k^1,\dots,\xi_k^d)\in U$, $k=1,\dots,L$, and coefficients
$\lambda_1,\dots,\lambda_L\in\mathbb C$ such that
\begin{equation}\label{cauchy_th}
\sup_{z\in M}
\left|
f(z)-\sum_{k=1}^L
\frac{\lambda_k}{\prod_{j=1}^d(\xi_k^j-z^j)}
\right|
<\varepsilon .
\end{equation}
Moreover, the approximation error admits algebraic convergence of
arbitrarily high order in $L$.
\end{theorem}

% \begin{remark}
% Theorem~\ref{Cauchy_approximation_theorem} corresponds to Theorem~1 in Xia~\cite{XNet1}.  
% The convergence order $p$ is determined by the smoothness properties of the target function $f$ and remains independent of the spatial dimension $d$. For analytic functions, which are infinitely differentiable, $p$ can be selected arbitrarily large, enabling exponential convergence rates. For general continuous functions, the density of analytic functions in the space of continuous functions ensures effective approximation through analytic intermediates. 
% \end{remark}
\begin{remark}[Precise meaning of the convergence order]
\textbf{The statement in Theorem~\ref{Cauchy_approximation_theorem} that the
approximation admits algebraic convergence of arbitrarily high order}
means that for any $p>0$, there exists a constant $C_p>0$, depending on $p$
and on $f$ but independent of $L$, such that
\[
\sup_{z\in M}
\left|
f(z)-\sum_{k=1}^L
\frac{\lambda_k}{\prod_{j=1}^d(\xi_k^j-z^j)}
\right|
\le C_p\,L^{-p}.
\]
In fact, when $f$ is analytic in a neighborhood of $M$, the approximation
error decays geometrically (i.e., exponentially) with respect to $L$,
from which the above algebraic bounds follow immediately. Here $p$ is an arbitrarily prescribed convergence order. For each fixed $p > 0$, the approximation error admits an $O(L^{-p})$ bound with a constant depending on p and the target function $f$, but independent of $L$. In the analytic case, this follows from the underlying exponential convergence.
\end{remark}

Motivated by the Cauchy approximation theorem and validated through extensive empirical studies across function approximation, PDE solving, and reinforcement learning applications, Xia et al. \cite{XNet1,XNet2} developed the XNet architecture with the following mathematical formulation:
\begin{equation}\label{XNet}
	\begin{aligned}
		Y_{\text{XNet}}^\theta(X) &= \text{Re}\left(\sum_{k=1}^L \frac{\alpha_k + i \beta_k}{\sum_{j=1}^d a_k^j x_j + c_k + i e_k}\right) \\
		&= \sum_{k=1}^L \left( \alpha_k \frac{\sum_{j=1}^d a_k^j x_j + c_k}{\left(\sum_{j=1}^d a_k^j x_j + c_k\right)^2 + e_k^2} + \beta_k \frac{e_k}{\left(\sum_{j=1}^d a_k^j x_j + c_k\right)^2 + e_k^2}\right),
	\end{aligned}
\end{equation}
where $X=(x_1,x_2,...,x_d)$ represents the $d$-dimensional input, $Y_{\text{XNet}}^\theta$ denotes the output, and $\alpha_k, \beta_k, a_k^j, c_k,$ and $e_k$ are trainable parameters.

The architectural simplicity of shallow XNet achieves an better balance between approximation capability and optimization cost. With $\mathcal{O}(L)$ parameters versus $\mathcal{O}(HL^2)$ for traditional FNNs, XNet dramatically reduces computational burden while maintaining superior approximation power. This efficiency is particularly valuable in Deep BSDE implementations requiring simultaneous training of multiple networks across temporal layers. Combined with dimension-independent convergence properties, XNet enables the extension of Deep BSDE methods to high-dimensional problems previously considered computationally intractable.

\section{Convergence of Deep BSDE method}\label{s3}

The Deep BSDE method transforms the PDE solving problem into a stochastic control framework. Consider the Euler discretization scheme:
\begin{equation}\label{system}
\small
	\left\{\begin{array}{l}
		{X}_0^\pi=\xi, \quad Y_0^\pi=\theta_{{u_0}}, \quad Z_0^{\theta,\pi}=\sigma^{\mathrm{T}}(t_0,\xi)\theta_{\nabla{u_0}}, \\
		{X}_{t_{n+1}}^\pi = {X}_{t_n}^\pi + \mu\left(t_n, {X}_{t_n}^\pi \right)\Delta t_n + 
		\sigma \left(t_n, {X}_{t_n}^\pi \right) \Delta W_n, \\
		Y_{t_{n+1}}^\pi = Y_{t_n}^\pi - f\left(t_n, {X}_{t_n}^\pi, Y_{t_n}^\pi, Z_{t_{n}}^{\theta,\pi}
        \right)\Delta t_n + \left[Z_{t_n}^{\theta,\pi}\right]^{\mathrm{T}}   \Delta W_n, \\
		Z_{t_{n}}^{\theta,\pi} = \sigma^{\mathrm{T}}\left(t_n, {X}_{t_n}^\pi \right) \nabla u^\theta_{t_n}(t_n,X^\pi_{t_n}),
	\end{array}\right.
\end{equation}
where $\nabla u^\theta_{t_n}$ represents the neural network approximation at time step $n$, which serves as the control strategy. The objective is to solve the optimization problem:
\begin{equation}\label{optimization}
	\mathop {\inf }\limits_{ {\theta _{{u_0}}},{\theta _{\nabla {u_0}}}  \in {\mathcal{N}_0},{\theta _i} \in {\mathcal{N}_i}} F(\theta) =\mathbb{E} \left[ {{\left| {g \left( {{X}_T^\pi} \right) - Y_{t_N}^\pi } \right|}^2}  \right],
\end{equation}
where $\mathcal{N}_0$ and $\mathcal{N}_i (0  \le  i \le N - 1)$ are parametric function spaces generated by neural networks.

In this section, we establish a comprehensive convergence theory for the Deep BSDE method, extending beyond the classical Lipschitz framework. The theoretical framework addresses three primary sources of computational error: approximation errors arising from temporal discretization and neural network representation, generalization errors from Monte Carlo sampling, and optimization errors due to network complexity and scale. Under the assumption that neural network approximation errors and generalization errors are sufficiently small relative to discretization errors, we establish convergence results for three fundamental categories of generators.

\subsection{Convergence Analysis for Lipschitz Generators}

The foundational convergence theory for the Deep BSDE method was established by Han et al. \cite{Convergence-of-the-deep-BSDE-method}, which requires the generator function $f$ to satisfy a global Lipschitz condition. Under standard regularity Assumptions \ref{assumption_1} and \ref{assumption_3} (detailed in Appendix A), this framework provides rigorous convergence guarantees for the Deep BSDE approximation scheme.

\begin{lemma}\label{lemma_2}
Under Assumptions \ref{assumption_1} and \ref{assumption_3}, there exists a constant $C_1$, independent of $h = \max_n \Delta t_n$, $d$, and $M$, such that 
\begin{equation}\label{lemma2}
	\begin{aligned}
		&\sup _{t_n \in[0, T]}\left(\mathbb{E}\left|X_{t_n}-{X}_{t_n}^\pi\right|^2+\mathbb{E}\left|Y_{t_n}-{Y}_{t_n}^\pi\right|^2\right)+\sum_{n=0}^{N-1}\int_{t_n}^{t_{n+1}} \mathbb{E}\left|Z_t-{Z}_{t_n}^\pi\right|^2 dt \\
		&\qquad\qquad\leq C_1\left[h+\mathbb{E}\left|g\left(X_T^\pi\right)-Y_T^\pi\right|^2\right].
	\end{aligned}
\end{equation}

\end{lemma}

\begin{lemma}\label{lemma_3}
Under Assumptions \ref{assumption_1} and \ref{assumption_3}, there exists a constant $C_2$, independent of $h$, $d$ and $M$, such that 
\begin{equation}\label{lemma3}
	\begin{aligned}
		&\inf_{\theta_{{u_0}},\theta_{\nabla{u_0}} \in \mathcal{N}_0, \phi_n \in \mathcal{N}_n} \mathbb{E}\left|g\left(X_T^\pi\right)-Y_T^\pi\right|^2 \\
		&\leq C_2\left\{h+\inf _{\theta_{{u_0}},\theta_{\nabla{u_0}} \in \mathcal{N}_0} \mathbb{E}\left| Y_0^\pi-\theta_{{u_0}} \right|^2 +  \mathbb{E}\left| Z_0^\pi-\theta_{\nabla{u_0}} \right|^2  h\right.\\
		&\left.\qquad + \inf _{\phi_n^\theta \in \mathcal{N}_n} \sum_{n=1}^{N-1} \mathbb{E}\left|\mathbb{E}\left[\tilde{Z}_{t_n} \mid X_{t_n}^\pi\right]-\phi_n^\theta \left(t_n, X_{t_n}^\pi\right)\right|^2 h\right\},
	\end{aligned}
\end{equation}	
where $\tilde{Z}_{t_n}=h^{-1}\mathbb{E}\left[\int_{t_n}^{t_{n+1}} Z_t \mathrm{d} t \mid \mathcal{F}_{t_n}\right]$. 
\end{lemma}

Lemmas \ref{lemma_2} and \ref{lemma_3} establish bounds on the absolute error of the Deep BSDE method. The detailed proofs can be found in Han et al. \cite{Convergence-of-the-deep-BSDE-method}. Given our primary interest in relative error calculations, we further derive bounds on the relative error:
\begin{equation}\label{th1}
	\begin{aligned}
	\mathbb{E}\left|\frac{u(0, \xi)-\theta_{{u_0}}}{u(0,\xi)}\right|^2
	 \leq &C\left\{ \frac{h}{u(0,\xi)^2} +\inf _{\theta_{{u_0}},\theta_{\nabla{u_0}} \in \mathcal{N}_0}  \mathbb{E}\left| Z_0^\pi-\theta_{\nabla{u_0}} \right|^2  \frac{h}{u(0,\xi)^2} \right.\\
	&\left. +  \inf _{\phi_n^\theta \in \mathcal{N}_n} \sum_{n=1}^{N-1} \mathbb{E}\left|\mathbb{E}\left[\tilde{Z}_{t_n} \mid X_{t_n}^\pi\right]-\phi_n^\theta\left(t_n, X_{t_n}^\pi\right)\right|^2 \frac{h}{u(0,\xi)^2}\right\},
	\end{aligned}
\end{equation}
where constant $C$ is independent of $h$, $d$ and $M$.

\begin{remark}\label{theorem_1} 
This analysis reveals that the approximation errors of the Deep BSDE method consist of two primary components: time discretization errors (standard in numerical methods) and neural network approximation errors. When $u(0,\xi)$ is sufficiently large, the time discretization error becomes negligible, and the overall approximation error is dominated by the network's approximation capability.
\end{remark}

While this theoretical framework is well-established for Lipschitz generators, many practically important PDEs violate the global Lipschitz condition. Notable examples include Allen-Cahn equations with cubic nonlinearity ($f(t,x,y,z) = y - y^3$) and Hamilton-Jacobi-Bellman equations with quadratic gradient growth ($f(t,x,y,z) = -\frac{1}{2}|z|^2$) from the original Deep BSDE work \cite{E2017}. Although these equations demonstrated numerical convergence in empirical studies, their theoretical convergence remained unexplained within the existing Lipschitz framework.
To address this theoretical gap, we extend the convergence analysis to these two important classes of non-Lipschitz equations. Through novel analytical techniques that exploit the special structural properties of these generators, we establish convergence results that provide the first rigorous theoretical justification for the empirical success observed in practice.

\subsection{Convergence Analysis for Allen-Cahn Type Equations}
Allen-Cahn equations with cubic nonlinearity represent a fundamental class of reaction-diffusion systems arising in phase field theory and materials science. Consider the $d$-dimensional Allen-Cahn equation:
\begin{equation}\label{eq:AllenCahn}
	\left\{\begin{array}{l}
		\frac{\partial u}{\partial t}(t, x)+\frac{1}{2}\text{Tr}[\sigma\sigma^T(t,x)D^2u(t,x)] + \mu(t,x) \cdot \nabla u(t,x)+u(t, x)-[u(t, x)]^3=0, \\
		\quad (t, x) \in[0, T) \times \mathbb{R}^d, \\
		u(T, x)=g(x), \quad x \in \mathbb{R}^d,
	\end{array}\right.
\end{equation}
where the generator function $f(t,x,y,z)=y-y^3$ exhibits the characteristic double-well potential structure. The cubic nonlinearity formally violates the global Lipschitz condition required in standard convergence theory.

The key insight for establishing convergence lies in exploiting the intrinsic boundedness properties induced by the double-well potential. The critical observation is that solutions to the BSDE \eqref{eq:BSDE-Allen-Cahn} exhibit natural bounds that prevent explosive growth, despite the super-linear nonlinearity.

\begin{lemma}[Boundedness Properties of Double-Well Dynamics]
\label{lem:L2boundY} 
Let $(Y_t,Z_t)_{t\in[0,T]}$ be the adapted solution to the BSDE
\begin{equation}\label{eq:BSDE-Allen-Cahn}
Y_t = Y_T + \int_t^T \bigl( Y_s - Y_s^3 \bigr)\,ds
      - \int_t^T Z_s\,dW_s ,
\end{equation}
where $Y_T\in L^2(\Omega)$.  
Then there exists a constant $C>0$, depending only on $T$ and
$\mathbb E[Y_T^2]$, such that
\[
\sup_{t\in[0,T]} \mathbb E\bigl[Y_t^2\bigr] \le C.
\]
In particular,
\[
\sup_{t\in[0,T]} \mathbb E\bigl[|Y_t|\bigr] < \infty .
\]
\end{lemma}

The proof follows from explicit integration using separation of variables (detailed in Appendix B). This boundedness result is crucial for controlling the cubic nonlinearity in the BSDE framework.

\begin{theorem} [Convergence for Allen-Cahn Equations] \label{th_conv_AC}
Consider the Allen-Cahn equation \eqref{eq:AllenCahn} with generator function $f(t,x,y,z) = y - y^3$. Under Assumptions \ref{assumption_1}, \ref{assumption_3}, the Deep BSDE method \eqref{system} converges. Specifically, there exists a constant $C_1,C_2 > 0$, independent of $h$, $d$, and $M$, such that 
\begin{equation}
\begin{aligned}
\sup _{0\le n \le N} \mathbb{E}\left|Y_{t_n}-Y_n^\pi\right|^2
	&\leq C_1\left[h+\mathbb{E}\left|g\left(X_T^\pi\right)-Y_T^\pi\right|^2\right]\\
\inf_{\theta_{{u_0}},\theta_{\nabla{u_0}} \in \mathcal{N}_0, \phi_n \in \mathcal{N}_n} \mathbb{E}\left|g\left(X_T^\pi\right)-Y_T^\pi\right|^2 
		&\leq C_2\left\{h+\inf _{\theta_{{u_0}},\theta_{\nabla{u_0}} \in \mathcal{N}_0}\mathbb{E}\left| Y_0^\pi-\theta_{{u_0}} \right|^2 +  \mathbb{E}\left| Z_0^\pi-Z_0^{\theta,\pi} \right|^2  h\right.\\
		&\left.\qquad + \inf _{\phi_n^\theta \in \mathcal{N}_n} \sum_{n=1}^{N-1} \mathbb{E}\left|\mathbb{E}\left[\tilde{Z}_{t_n} \mid X_{t_n}^\pi\right]-Z_n^{\theta,\pi}\right|^2 h\right\}.
	\end{aligned}
\end{equation}
\end{theorem}

\begin{proof}
From the BSDE representation (equation \ref{Yt}) and Lemma \ref{lem:L2boundY}, we know that $\mathbb{E}|Y_t|$ is bounded. Specifically, for any two solutions $Y_{1,t}$ and $Y_{2,t}$, we have:
\begin{equation}
\begin{aligned}
&\mathbb{E}|f(t,X_t,Y_{1,t},Z_t) - f(t,X_t,Y_{2,t},Z_t)| \\
= &\mathbb{E}|Y_{1,t} - Y_{2,t}||1 - Y_{1,t}^2 - Y_{1,t}Y_{2,t} - Y_{2,t}^2| \\
\leq &K \mathbb{E}|Y_{1,t} - Y_{2,t}|,
\end{aligned}
\end{equation}
where the constant $K$ depends on the uniform bounds of $Y_t$. This local Lipschitz property enables application of the standard convergence framework. 
\end{proof}

\begin{remark}
This result resolves the theoretical gap for Allen-Cahn equations, providing rigorous justification for the numerical convergence observed in the original Deep BSDE work \cite{E2017} and our numerical experiments in Section \ref{s5}. 
\end{remark}

\subsection{Convergence Analysis for HJB Type Equations}
Hamilton-Jacobi-Bellman (HJB) equations represent a fundamental class of nonlinear PDEs arising in stochastic optimal control and mathematical finance. These equations present significant theoretical challenges due to their quadratic gradient dependence, which violates standard Lipschitz conditions required for classical convergence analysis. Consider the $d$-dimensional HJB equation:
\begin{equation}\label{eq:HJB}
\left\{\begin{array}{l}
\frac{\partial u}{\partial t}(t, x) + \frac{1}{2}\text{Tr}[\sigma\sigma^T(t,x)D^2u(t,x)] + \mu(t,x) \cdot \nabla u(t,x) - \frac{1}{2}|\sigma^T(t,x)\nabla u(t,x)|^2 = 0, \\
u(T, x) = g(x), \quad x \in \mathbb{R}^d,
\end{array}\right.
\end{equation}
where the generator function is $f(t,x,y,z) = -\frac{1}{2}|z|^2$. The quadratic dependence on $z$ violates the standard Lipschitz condition in $H_2$ in Assumption \ref{assumption_1}, presenting a significant theoretical challenge.

The convergence analysis for HJB equations requires a fundamentally different approach than the Allen-Cahn case. The quadratic growth of the generator with respect to the control variable $z$ can lead to explosive behavior, necessitating sophisticated truncation and regularization techniques. Our proof strategy establishes convergence through a systematic analysis of four interconnected stochastic systems.

We begin by introducing a truncated BSDE system where the quadratic generator is regularized through projection onto a ball of radius $B$:
\begin{equation}\label{eq:truncated_system}
\left\{ {\begin{aligned}
& {X_t = \xi + \int_0^t \mu(s,X_s)\,ds + \int_0^t \sigma(s,X_s)\,dW_s}, \\
& { Y_t^B = g(X_T) + \int_t^T f^B(s,X_s,Y_s^B,Z_s^B)\,ds - \int_t^T Z_s^B\,dW_s},
\end{aligned}} \right.
\end{equation}
where $f^B(\cdot,\cdot,\cdot,z) = f(\cdot,\cdot,\cdot,\varphi^B(z))$ 
and $\varphi^B$ is the projection on the centered Euclidean ball
of radius $\rho B$ with $\rho >0$ chosen such that $f^B$ is B-Lipschitz-continuous with respect to $z$.

The truncation error between the original and truncated systems is controlled by the following result from the BSDE literature:
\begin{lemma}[Truncation Error Control; Theorem 6.2 in \cite{imkeller2010path}, Remark 5.5 in \cite{richou2012markovian}]
\label{lemma:Truncation_Convergence_brief}
Under the \textbf{$H_1$}, \textbf{$H_3$} in Assumption \ref{assumption_1} and Assumption \ref{ass:quadratic}, for any $p \geq 1$ and $\beta \geq 1$, there exist positive constants $C_p$ and $D_\beta$ such that
\[
\mathbb{E}\!\left[\sup_{t \in [0,T]} |Y_t^B - Y_t|^{2p}\right] + \mathbb{E}\!\left[\Big(\int_{0}^{T} |Z_s^B - Z_s|^2 \, ds \Big)^p\right] \leq C_p D_\beta B^{-\tfrac{\beta}{2\bar{q}}}.
\]
\end{lemma}
This lemma establishes that the truncated system \eqref{eq:truncated_system} converges to the original BSDE system \eqref{BSDE} as $B \to \infty$. The constants $C_p$ and $D_\beta$, as well as the Hölder conjugate $\bar{q}$, are characterized in Theorem 6.2 of Imkeller \cite{imkeller2010path} and Remark 5.5 of Richou \cite{richou2012markovian}.

Having established truncation error control, our primary task reduces to proving convergence of the Deep BSDE system \eqref{system} to the truncated system \eqref{eq:truncated_system}. This requires a reformulation within the Bouchard–Touzi–Zhang (BTZ) framework. 
Right multiplying $(\Delta W_i)^{T}$ on both sides of \eqref{system}, and taking the expectation conditional expectation $\mathbb{E}[\cdot\,|\,\mathcal{F}_n]$ again, we obtain
\begin{align*}
\mathbb{E}\!\left[Y_{n+1}^\pi (\Delta W_n)^T \big| \mathcal{F}_n \right] 
  = h \, Z_n^{\theta,\pi}.
\end{align*}
The above observation motivates us to consider the following BTZ scheme \cite{bouchard2004discrete,zhang2004numerical}:
\begin{equation}\label{eq:BTZ_system}
\begin{aligned}
\overline{X}_0^\pi &= \xi, \\
\overline{X}_{n+1}^\pi &= \overline{X}_n^\pi + \mu(t_n, \overline{X}_n^\pi) h + \sigma(t_n, \overline{X}_n^\pi) \Delta W_n, \\
\overline{Y}_n^\pi &= \mathbb{E}_n[\overline{Y}_{n+1}^\pi + f^B(t_n, \overline{X}_n^\pi, \overline{Y}_n^\pi, \overline{Z}_n^\pi) h], \\
\overline{Z}_n^\pi &= \frac{1}{h} \mathbb{E}_n[\overline{Y}_{n+1}^\pi \Delta W_n].
\end{aligned}
\end{equation}
The truncated system \eqref{eq:truncated_system} also admits a perturbed system \eqref{eq:BTZ_system} representation.
\begin{equation}\label{eq:perturbed_BTZ}
\begin{aligned}
 X_{n+1}^\pi &= X_n^\pi + \mu(t_n, X_n^\pi) h + \sigma(t_n, X_n^\pi) \Delta W_n + \Upsilon_n^X, \\
\widetilde{Y}_n^\pi &= \mathbb{E}_n[\widetilde{Y}_{n+1}^\pi + h f^B(t_n, X_n^\pi, \widetilde{Y}_n^\pi, \widetilde{Z}_n^\pi)] + \Upsilon_n^Y, \\
\widetilde{Z}_n^\pi &= \frac{1}{h} \mathbb{E}_n[\widetilde{Y}_{n+1}^\pi \Delta W_n],
\end{aligned}
\end{equation}
where $\widetilde{Y}_n^{\pi} = Y_{t_n}^B$, and the perturbation term satisfies:
\begin{equation}\label{eq:Upsilon}
\begin{aligned}
\Upsilon^X_n &= \int_{t_n}^{t_{n+1}} \mu(s,{X}_{s}) - \mu(t_n, X^\pi_n) \, ds 
+ \int_{t_n}^{t_{n+1}} \sigma(s,\widetilde{X}_{s}) - \sigma(t_n, X^\pi_n)\, dW_s,\\
\Upsilon_n^Y &= \mathbb{E}_n\left[\int_{t_n}^{t_{n+1}} \big(f^B(s,X_s,Y_s^B,Z_s^B) - f^B(t_n,X_n^\pi,Y_{t_n}^B,\widetilde{Z}_n^\pi)\big) ds\right].
\end{aligned}
\end{equation}

We prove that the Deep BSDE system converges to the truncated BSDE system within the BTZ framework by Theorem \ref{Th:estimation_HJB} and Theorem \ref{th:DBSDE_conv_PBTZ_brief}.

{
\begin{theorem}\label{Th:estimation_HJB}
Let \(\bigl(X_n^\pi, \widetilde{Y}_n^\pi, \widetilde{Z}_n^\pi \bigr)\) denote the solution to the truncated BSDE system \eqref{eq:perturbed_BTZ}, \(\bigl(X_{n}^\pi, Y_{n}^\pi, Z_{t_n}^{\theta,\pi}\bigr)\) denote the solution of the Deep BSDE (DBSDE) scheme \eqref{system}. Under the \textbf{$H_1$}, \textbf{$H_3$} in Assumption \ref{assumption_1}, Assumption \ref{assumption_3} and Assumption \ref{ass:quadratic},
there exists a constant \(C\), such that 
\[
\sup_{0\le n\le N}\mathbb E|\widetilde Y_n^\pi - Y^\pi_n|^2
\le C (h^{\frac{1}{2}}+\mathbb E|\widetilde Y_N^\pi-Y_N^\pi|^2 ).
\]
\end{theorem}}

\begin{theorem}\label{th:DBSDE_conv_PBTZ_brief}
Under the \textbf{$H_1$}, \textbf{$H_3$} in Assumption \ref{assumption_1}, Assumption \ref{assumption_3} and Assumption \ref{ass:quadratic}, the Deep BSDE system converges to the truncated BSDE system with rate
\begin{equation}
\mathbb{E}|\widetilde Y_N^\pi - Y^\pi_N| \leq e^{C_f T} (\mathbb{E}_0\!\big| Y_0 - \theta_{u_0} \big| + \sum_{n=0}^{N-1}\mathbb{E}_n\!\big| \left(Z^{\theta,\pi}_{n} - \widetilde{Z}_n^\pi\right)^2 \big|h) + O(h^{\frac{1}{2}}),
\end{equation}
where $\widetilde{Z}_n^\pi = \frac{1}{h} \mathbb{E}_n[Y_{t_{n+1}}^B \Delta W_n]$ represents the target approximation for the neural networks.
\end{theorem}

The proof of Theorem \ref{th:DBSDE_conv_PBTZ_brief} requires careful analysis of the perturbation terms arising from the continuous-to-discrete approximation. The key technical challenge lies in controlling the quadratic nonlinearity while maintaining the martingale structure. The complete proof is provided in Appendix C.
Combining the results of Lemma \ref{lemma:Truncation_Convergence_brief} and Theorem \ref{th:DBSDE_conv_PBTZ_brief} yields our main convergence theorem:

{
\begin{theorem}[Convergence for HJB Equations] \label{th_conv_HJB}
Under the \textbf{$H_1$}, \textbf{$H_3$} in Assumption \ref{assumption_1}, Assumption \ref{assumption_3} and Assumption \ref{ass:quadratic}, the Deep BSDE method converges for HJB equations with quadratic generators. Specifically, there exists a constant $C_1',C_2' > 0$ such that
\begin{equation}\label{eq:HJB_convergence}
\begin{aligned}
\sup _{0\le n \le N}\mathbb E|Y_{t_n} - Y_n^\pi|^2 &\le C_1' (h^{1/2}+\mathbb E|g(X_T^\pi)-Y_N^\pi|^2 + C_p D_\beta B^{-\tfrac{\beta}{2\bar{q}}})\\
\inf_{\theta_{{u_0}},\theta_{\nabla{u_0}} \in \mathcal{N}_0, \phi_n \in \mathcal{N}_n} \mathbb{E}\left|{g(X_T^\pi)-Y^\pi_N}\right|  &\leq C_2'\left(h^{\frac{1}{2}} + \inf _{\theta_{{u_0}},\theta_{\nabla{u_0}} \in \mathcal{N}_0}  \mathbb{E}\left|Y_{0} - \theta_{u_0}\right|^{2} +  \mathbb{E}\left| \widetilde Z_0^\pi-Z_0^{\theta,\pi} \right|^2 h \right.\\
& \left. \qquad  +  \inf _{\phi_n^\theta \in \mathcal{N}_n} \sum_{n=1}^{N-1} \mathbb{E}\left|\widetilde{Z}_n^\pi-Z_n^{\theta,\pi} \right|^2 h \right)
+ \sqrt {C_p D_{\beta}} B^{-\frac{\beta}{4q}}.
\end{aligned}
\end{equation}
\end{theorem}}

\begin{remark}
The convergence analysis reveals a fundamental difference in the target approximation requirements between Lipschitz and quadratic generators. For Lipschitz generators, neural networks must approximate $h^{-1} \mathbb{E} \left[ \int_{t_n}^{t_{n+1}} Z_t \, dt \;\middle|\; \mathcal{F}_{t_n} \right]$, which represents the time-averaged $Z$ process. For quadratic generators, the target becomes $\frac{1}{h} \mathbb{E}_n[Y_{t_{n+1}}^B \Delta W_n]$, which captures the martingale structure of the truncated system. 
\end{remark}

\begin{remark}
This theoretical advancement significantly extends the applicability of Deep BSDE methods to nonlinear PDEs arising in portfolio optimization, risk management, and optimal control problems. The rigorous convergence guarantees provide theoretical foundation for the empirical success observed in computational finance applications \cite{davey2024deep, han2018solving,E2017}.
\end{remark}

\subsection{Generalization Errors and Optimization Errors for Deep BSDE method}
In the Deep BSDE method, the loss function is defined as the expectation of matching the terminal condition (\ref{optimization}). However, in practical computations, the exact expectation is not directly computed. Instead, we approximate it using the loss function \eqref{BSDE-LOSS}. This approximation results in a generalization error,
\begin{equation}\label{Generalization_Error}
	\begin{aligned}
	\text{Generalization Error}  &	=\mathbb{E} {{\left| {g \left( {{X}_T} \right) - Y_{t_N} } \right|}^2}  - \frac{1}{M}\sum\limits_{m = 1}^M {{{\left| {g\left( {{X_T^m}} \right) -  Y_{{t_N}}^m } \right|}^2}}\\
	& =\mathcal{L}  \left( g \right) - \mathcal{L} \left( g_M \right).
	\end{aligned}
\end{equation}

\begin{lemma}
For Monte Carlo methods $(\ref{system})$ that use i.i.d. samples, the convergence rate of the mean generalization error is
\begin{equation}\label{generalization}
	\mathbb{E}\left[\mathcal{L}(g)-\mathcal{L}(g_M)\right]=O(M^{-1/2+\varepsilon}),
\end{equation}
where $\epsilon$ is arbitrarily small constant, function $g$ satisfes the boundary growth condition $(\ref{boundary_growth})$ for some small constants $(B_i)_{i=1}^d$.  
This result follows directly from the work of Xiao et al. {\cite{Xiao}}.
\end{lemma}

\begin{definition}
(Boundary growth condition). Let $d\in\mathbb{N}$, suppose $\mathcal{G}$ is a class of realvalued functions defined on $(0,1)^d.$ We say that $\mathcal{G}$ satisfes the boundary growth condition with constants $(B_i)_i=1^d$ if there exists $B\in(0,\infty)$ such that for every $g\in\mathcal{G}$, every subset $v\subseteq\{1,\cdots,d\}$ and every $u=(u_1,\ldots,u_d)\in(0,1)^d$ it holds that
\begin{equation}\label{boundary_growth}
\left|\left(\prod_{i\in v}\partial/\partial x_i\right)g(u)\right|\leq B\prod_{i=1}^d[\min(u_i,1-u_i)]^{-B_i-\mathbf{1}\{i\in v\}},
\end{equation}
where $\mathbf{1}\{ \cdot \} $ is an indicator function.
\end{definition}

When using general Monte Carlo methods to sample trajectories (\ref{Xt}), a significant number of samples is required to reduce the generalization error to an adequately small level. Therefore, it is necessary to introduce techniques such as importance sampling {\cite{Importance_sampling}}, quasi-Monte Carlo methods {\cite{QMC}}, Gibbs sampling {\cite{Gibbs}}, and other advanced sampling techniques. These methods enable the reduction of generalization errors with a smaller sampling cost.

%The optimization problem involves minimizing the defined loss function \ref{BSDE-LOSS} by adjusting the network parameters. Increasing the number of parameters in a neural network typically leads to higher optimization errors. The primary reason for this effect is that a larger parameter space requires a more exhaustive search, making it more difficult for the optimization algorithm to converge to the global optimum, ultimately increasing the optimization errors {\cite{Parameter_optimization,Shen_Z}}. This phenomenon is commonly referred to as the challenge of escaping saddle points. Generally, simpler networks are associated with lower optimization errors.

The optimization problem aims to minimize the loss function \eqref{BSDE-LOSS} by adjusting network parameters. Smaller network architectures define more restricted function spaces, which reduce the parameter search space and promote faster convergence to high-quality optima. The resulting lower-dimensional optimization landscape enhances algorithmic efficiency, leading to reduced optimization error \cite{Parameter_optimization,Shen_Z}. This advantage is particularly important in the Deep BSDE framework, where multiple networks are trained simultaneously across time discretization steps. 

\begin{remark}
While the theoretical framework provides convergence guarantees, practical implementation depends heavily on controlling approximation and optimization errors. Generalization errors can be minimized by increasing the number of samples, but the choice of network architecture plays a pivotal role in controlling both approximation and optimization errors, which ultimately determine whether the theoretically predicted convergence behavior can be observed in computational experiments. The limitations of standard feedforward neural networks (FNNs) in achieving the required target approximation accuracy are evident in the experimental results. By incorporating the XNet architecture into the Deep BSDE framework, these shortcomings are effectively mitigated, allowing the method to achieve significantly smaller approximation and optimization errors under a constrained computational budget. This enhancement bridges the gap between theoretical convergence guarantees and practical implementation, revealing a more discernible convergence rate, as demonstrated in the numerical studies.
\end{remark}

\section{Discrete time models}\label{s4}
In the Deep BSDE method, we fully connect the $N-1$ temporal steps of the neural networks and train them jointly. At each time step, two options are available for the neural network architecture. 
For the feedforward neural networks (FNNs), the architecture comprises one input layer (d-dimensional), two hidden layers (each with d+10 dimensions), and one output layer (d-dimensional).
The XNet, on the other hand, consists of three components, including one input layer ($d$-dimensional), a hidden layer comprising $d$ basis functions, and one output layer ($d$-dimensional). 
Through the following two numerical examples, we demonstrate how the Deep BSDE method achieves enhanced accuracy and computational efficiency by replacing feedforward neural networks with XNet. Note that the numerical results obtained by the Deep BSDE method using feedforward neural networks are based on the code provided by E et al. {\cite{E2017}}.

\subsection{Allen-Cahn Equation}
In this subsection, the Deep BSDE method is tested for solving the 100-dimensional Allen-Cahn partial differential equation (PDE) (\ref{AllenCahneq}) using both the FNNs and XNet. With reference to the general form of the semilinear parabolic equation (\ref{bxxpwxpde}), we set $\alpha=1$, $f(y,z) = y - y^3$, and $g(x) = \left[2 + \frac{2}{5} |x|_{\mathbb{R}^d}^2\right]^{-1}$. The PDE is represented as follows,
\begin{equation}\label{AllenCahneq}
	\left\{\begin{array}{l}
		\frac{\partial u}{\partial t}(t, x)+\left(\Delta_x u\right)(t, x)+u(t, x)-[u(t, x)]^3=0,(t, x) \in[0, T) \times \mathbb{R}^d, \\
		u(T, x)=g(x), x \in \mathbb{R}^d,
	\end{array}\right.
\end{equation}
where the spatial dimension is $d=100$ and the terminal time $T=\frac{3}{10}$. Using the branching diffusion method {\cite{FENZHI1,FENZHI2,FENZHI3}}, a reference value for the exact solution is obtained, $u(0, \xi)=u(0,0, \ldots, 0)\approx 0.052802$. 
The Deep BSDE method is implemented by the FNNs and the XNet with setting the time step number to $N=20, 30, 40$, $80$, and conducting five independent runs for each configuration. During the training process, the numerical solution tends to stabilize after approximately 5000 iterations. Therefore, the average of the results from iterations 5000 to 10000 is taken as the computed value function. The results are presented in Table \ref{dt_allen_cahn_results}. Note that in the following figures, we use the term "Two-Layer Net" to specifically refer to the feedforward neural networks (FNNs) with two hidden layers as described above.

\begin{figure}[ht]
	\centering
	\begin{minipage}[t]{0.493\linewidth}
		\centering
		\includegraphics[width=\textwidth]{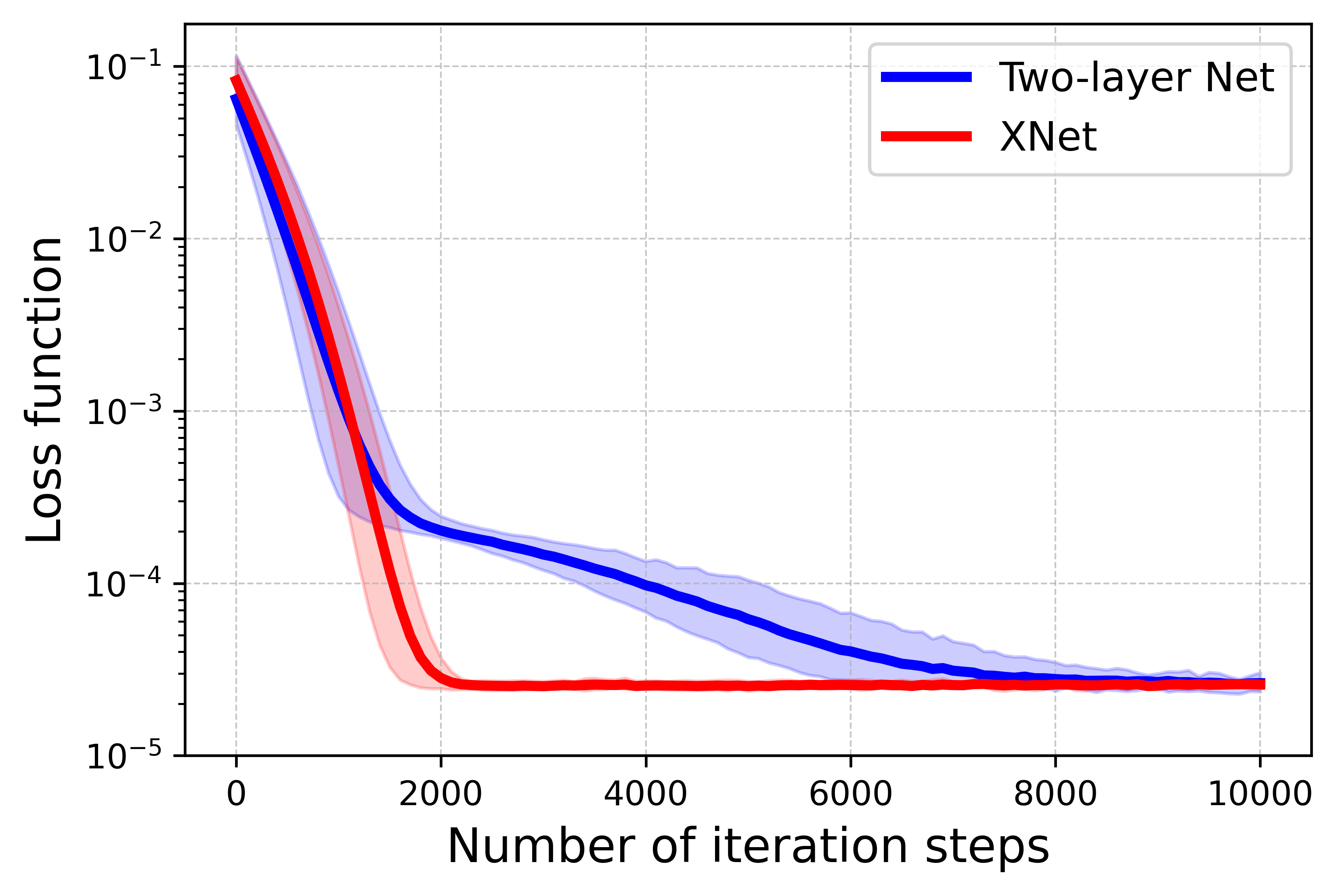}
		%\centerline{(a) Relative ${L^1}$-approximation error}
		\label{fig_dt_allencahn_20:first}
	\end{minipage}%
	\hfill
	\begin{minipage}[t]{0.5\linewidth}
		\centering
		\includegraphics[width=\textwidth]{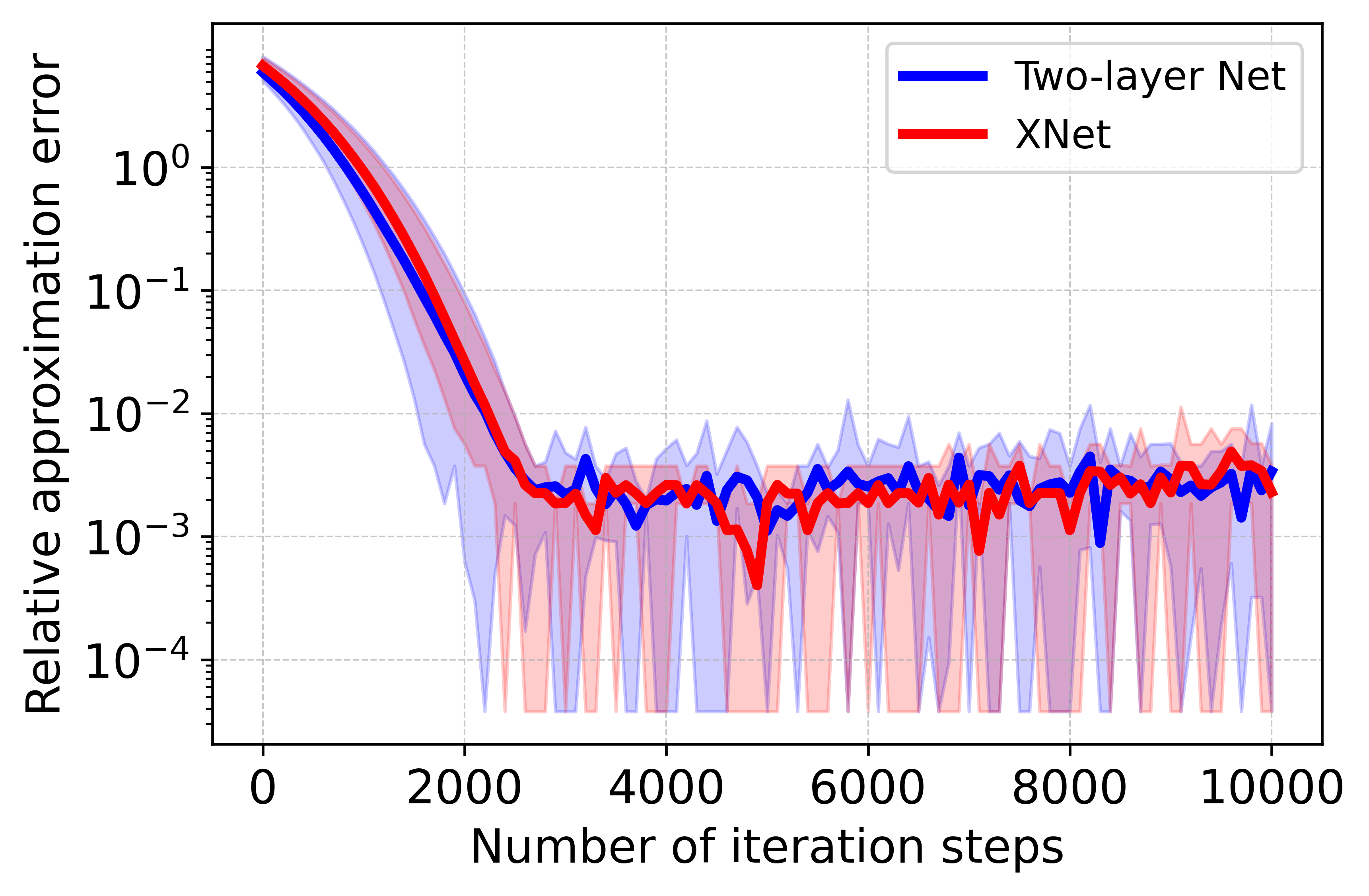}
		%\centerline{(b) Relative ${L^1}$-approximation error}
		\label{fig_dt_allencahn_20:second}
	\end{minipage}
	\vspace{-0.7cm}
	\centering
	\caption{Comparison of Two Network Architectures for Solving the Allen-Cahn Equation under 20-step-time Discretization}
	\label{fig_dt_allencahn1}
\end{figure}

Under a 20-time-step discretization, as shown in Figure \ref{fig_dt_allencahn1}, it is observed that switching to the XNet results in a faster decrease of the loss function and an increase in computational speed. However, accuracy improvement is not significant. This can be explained by Equation (\ref{allenth}), where the small value function indicates that the approximation error is dominated by the time discretization rather than the network approximation error,
\begin{equation}\label{allenth}
	\begin{aligned}
		\mathbb{E}\left|\frac{{u(0, \xi)-\theta_{{u_0}}}}{{u(0,\xi)}}  \right|^2\
		\leq &C\left\{ \frac{h}{0.052802^2} +\inf _{\theta_{{u_0}},\theta_{\nabla{u_0}} \in \mathcal{N}_0}  \mathbb{E}\left| Z_0-\theta_{\nabla{u_0}} \right|^2  \frac{h}{0.052802^2} \right.\\
		&\left. +  \inf _{\phi_n \in \mathcal{N}_n} \sum_{i=0}^{N-1} \mathbb{E}\left|\mathbb{E}\left[\tilde{Z}_{t_n} \mid X_{t_n}^\pi, Y_{t_n}^\pi\right]-\phi_n\left(X_{t_n}^\pi, Y_{t_n}^\pi\right)\right|^2 \frac{h}{0.052802^2}\right\}.
	\end{aligned}
\end{equation}

\begin{figure*}
	\centering
	\begin{minipage}[t]{0.5\linewidth}
		\centering
		\includegraphics[width=\textwidth]{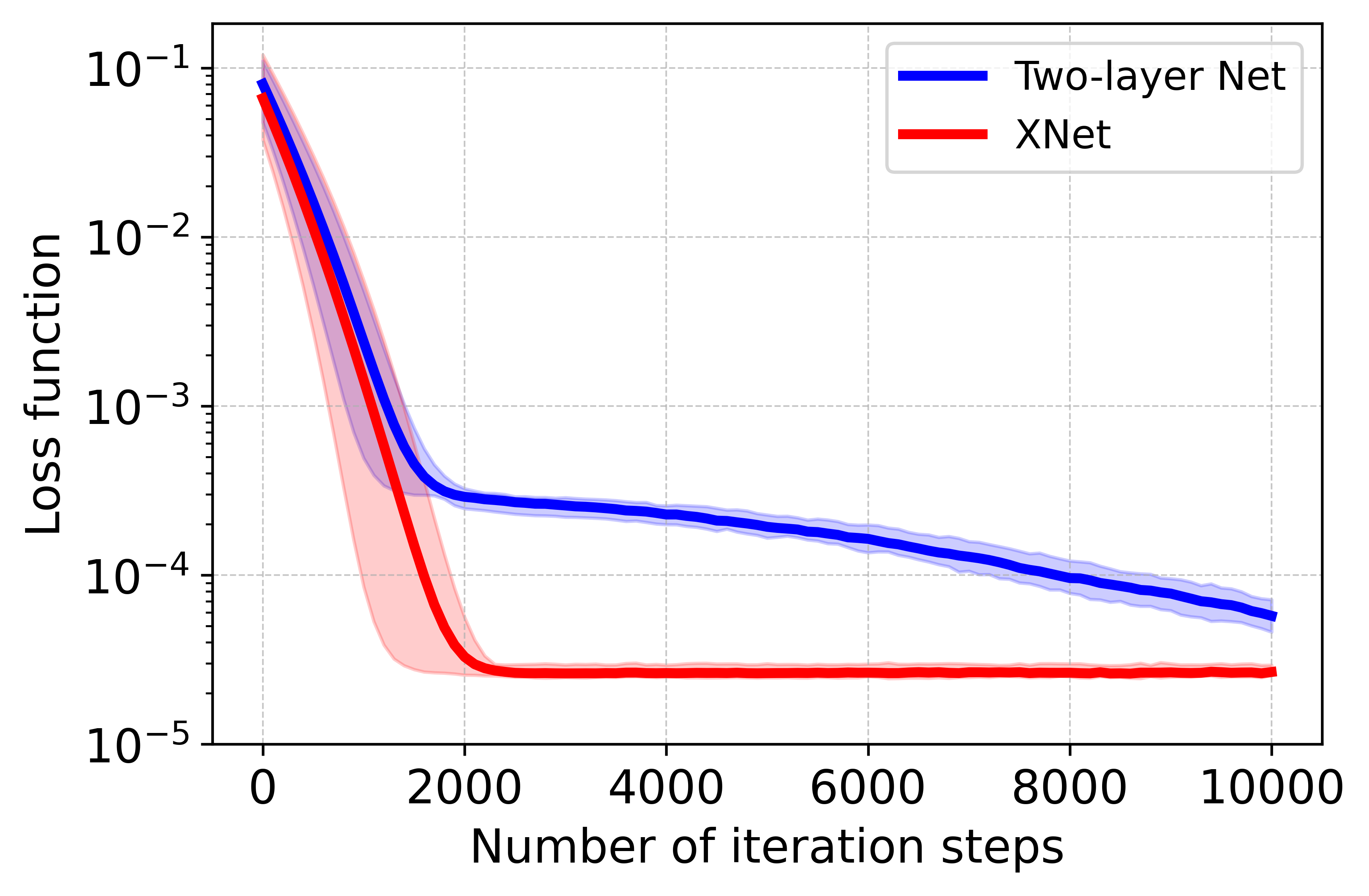}
		%\centerline{(a) Relative ${L^1}$-approximation error}
		\label{fig_dt_allencahn_20:first}
	\end{minipage}%
	\hfill
	\begin{minipage}[t]{0.5\linewidth}
		\centering
		\includegraphics[width=\textwidth]{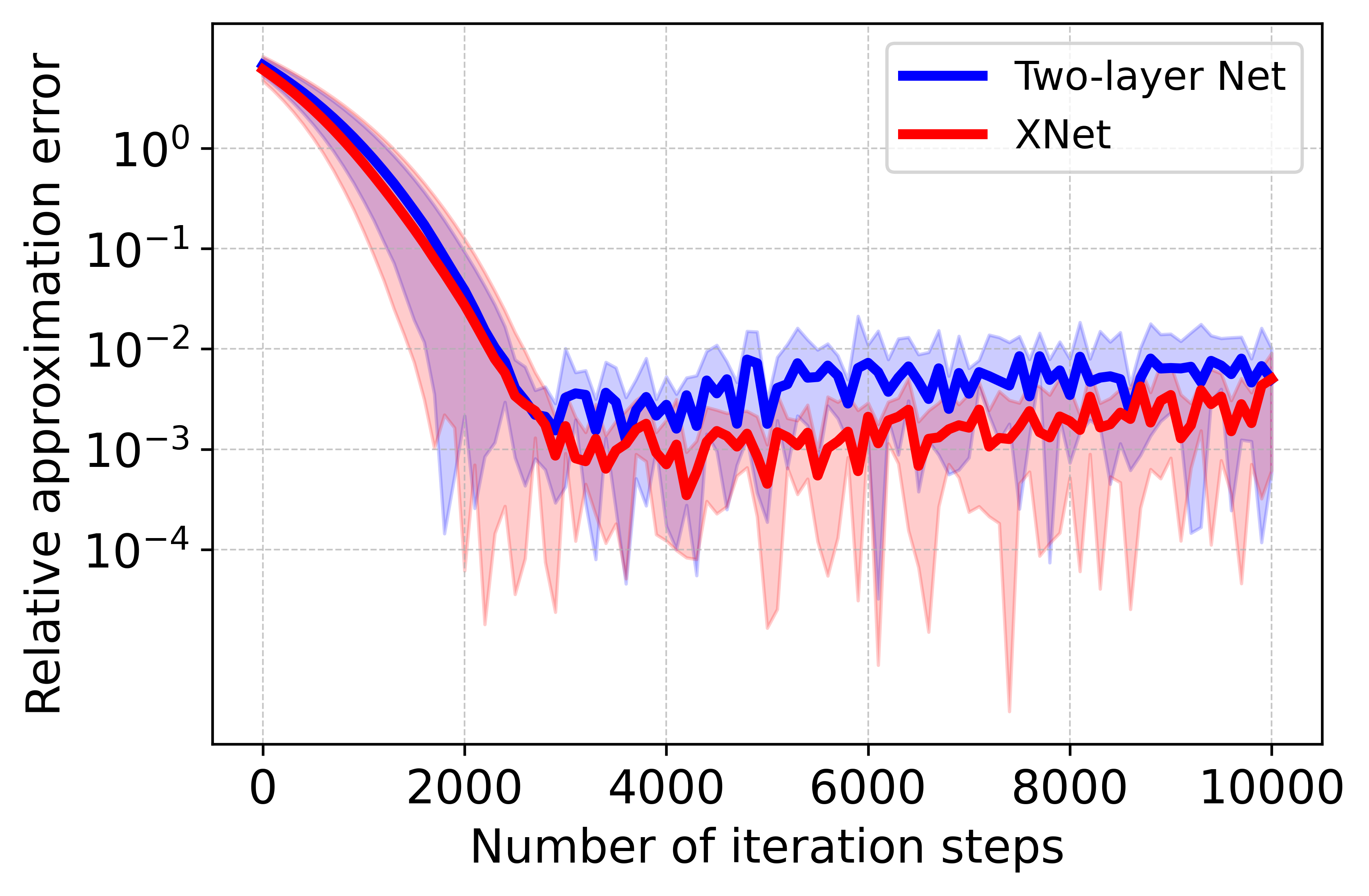}
		%\centerline{(b) Relative ${L^1}$-approximation error}
		\label{fig_dt_allencahn_20:second}
	\end{minipage}
	\vspace{-0.7cm}
	\centering
	\caption{Comparison of Two Network Architectures for Solving the Allen-Cahn Equation under 80-step-time discretization}
	\label{fig_dt_allencahn2}
\end{figure*}

As shown in Table 1, as the temporal discretization step size increases and h decreases, the approximation error associated with temporal discretization reduces. In this process, it is observed that the XNet, with its superior approximation capabilities, yields higher accuracy. However, the implementation of the FNNs fails to result in further improvement in the computational results. This phenomenon can be attributed to XNet providing smaller neural network approximation errors and optimization errors in the Deep BSDE method.
As shown in Figure \ref{fig_dt_allencahn2}, when the temporal discretization reaches 80 time steps, the results indicate that the Deep BSDE method implemented with the XNet results in a faster decrease in the loss function, higher computational efficiency, and greater accuracy. 
%In fact, we have already achieved very good results. The reason we do not further refine the time discretization is due to concerns that a significant increase in network parameters would lead to optimization errors.

\begin{table}[htbp]
	\centering
	\caption{Numerical Results for Allen-Cahn Equation}
	\resizebox{\textwidth}{!}{%
		\begin{tabular}{ccccccccc}
			\hline \hline
			& \multicolumn{4}{c}{XNet} & \multicolumn{4}{c}{Feedforward neural networks} \\
			\midrule
			Time steps & Runtime (s) & Value function & Relative Error  & Std. Deviation 
			& Runtime (s) & Value function & Relative Error  & Std. Deviation  \\
			\midrule
			20 & 72 & 5.2899e-02 & 1.8337e-03 & 5.7723e-05 & 83&5.2887e-02 & 1.6154e-03 & 6.1286e-05 \\
			30 & 112& 5.2877e-02 & 1.4209e-03 & 7.4003e-05 & 138 & 5.2875e-02 & 1.3807e-03 & 9.6353e-05 \\
			40 & 196 & 5.2846e-02 & 8.3214e-04 & 6.5463e-05 & 260 &5.2849e-02 & 8.8848e-04 & 1.5200e-04 \\
			80 & 464 &5.2820e-02 & 3.4374e-04 & 4.8460e-05 & 691&5.2867e-02 & 1.2309e-03 & 1.7797e-04 \\
			\hline \hline
		\end{tabular}%
	}
	\label{dt_allen_cahn_results}
\end{table}

\subsection{Pricing of European financial derivatives with different interest rates for borrowing and lending (PricingDiffrate) equation}
In this example, we consider a specialized nonlinear Black-Scholes equation. This equation models the pricing problem of European financial derivatives in a financial market where the risk-free bank account utilized for hedging purposes exhibits differential interest rates for borrowing and lending {\cite{Bergman4}}. Referring to the general form of the semi-linear parabolic PDE (\ref{bxxpwxpde}), we set $\bar{\mu}=0.06$, $\mu(t,x)=\bar{\mu}x$, $\bar{\sigma}=0.2$, $\sigma(t,x)=\bar{\sigma}x$. We assume for all $s,t\in[0,T]$, $x=(x_1,\ldots,x_d)\in\mathbb{R}^d$, $y\in\mathbb{R}$, and $z\in\mathbb{R}^{d}$, with $d=100$, $T=1/2$, and $\xi=(100,100,\ldots,100)\in\mathbb{R}^d$. Additionally, a terminal condition $g(x)$ and a non-linear term $f(t,x,y,z)$ are specified for the equation:
\begin{equation}\label{EPDFeqg}
	\begin{array}{l}
		g(x) = \max \left\{ {\left[ {{{\max }_{1 \le i \le 100}}{x_i}} \right] - 120,0} \right\} - 2\max \left\{ {\left[ {{{\max }_{1 \le i \le 100}}{x_i}} \right] - 150,0} \right\},
	\end{array}
\end{equation}
\begin{equation}\label{EPDFeqf}
	\begin{array}{l}
		f(t,x,y,z) =  - {R^l}y - \frac{{\left( {\bar \mu  - {R^l}} \right)}}{{\bar \sigma }}\sum\limits_{i = 1}^d {{z_i}}  + \left( {{R^b} - {R^l}} \right)\max \left\{ {0,\left[ {\frac{1}{{\bar \sigma }}\sum\limits_{i = 1}^d {{z_i}} } \right] - y} \right\},
	\end{array}
\end{equation}
where $R^l=0.04$, $R^b=0.06$. The equation can thus be represented on the domain $t \in [0, T)$ and $x \in \mathbb{R}^d$: 
\begin{equation}\label{EPDFeq}
	\begin{aligned}
		\frac{\partial u}{\partial t}(t,x)
		&+\frac{\bar{\sigma}^{2}}{2}\sum_{i=1}^{d}|x_{i}|^{2} \frac{\partial^{2}u}{\partial x_{i}^{2}}(t,x) +\bar{\mu}\sum_{i=1}^{d}x_{i} \frac{\partial u}{\partial x_{i}}(t,x)\\
		&+f\big(t,x,u(t,x),\bar{\sigma}\operatorname{diag}_{\mathbb{R}^{d\times d}}(x_{1},\ldots,x_{d})(\nabla_{x}u)(t,x)\big)=0.
	\end{aligned}
\end{equation}
The reference solution to equation \eqref{EPDFeq} is obtained using the Multilevel-Picard approximation method, which yields a value of $21.299$.
The Deep BSDE method is implemented using both FNNs and XNet architectures with temporal discretization parameters $N=20, 30, 40, 80$, conducting five independent runs for each configuration. The final computed value function represents the average over iterations 5000 to 10000. The results are presented in Table \ref{dt_diffrate_results}.

\begin{figure*}
	\centering
	\begin{minipage}[t]{0.5\linewidth}
		\centering
		\includegraphics[width=\textwidth]{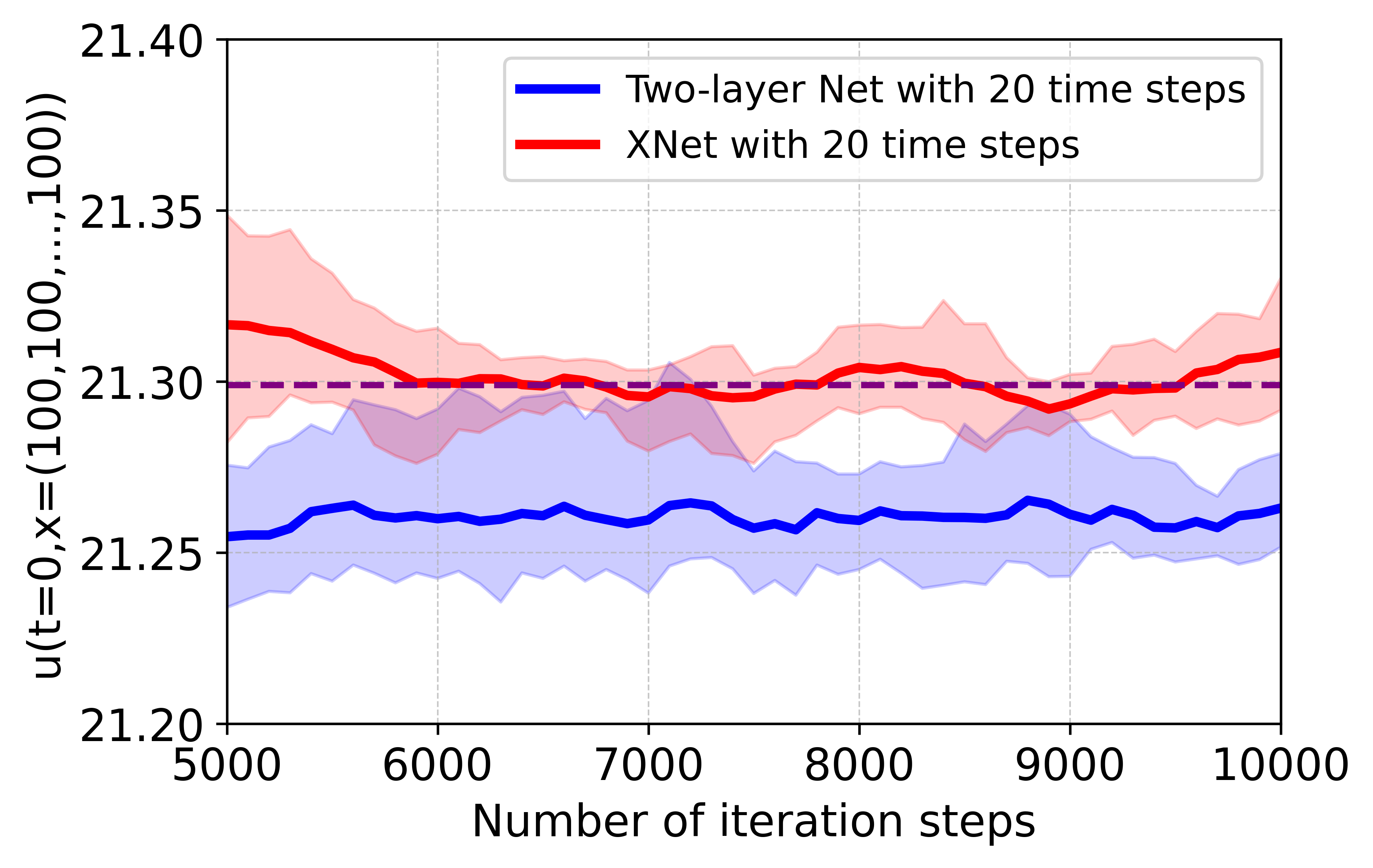}
		%\centerline{(a) Relative ${L^1}$-approximation error}
		\label{fig_dt_diffrate_80:first}
	\end{minipage}%
	\hfill
	\begin{minipage}[t]{0.5\linewidth}
		\centering
		\includegraphics[width=\textwidth]{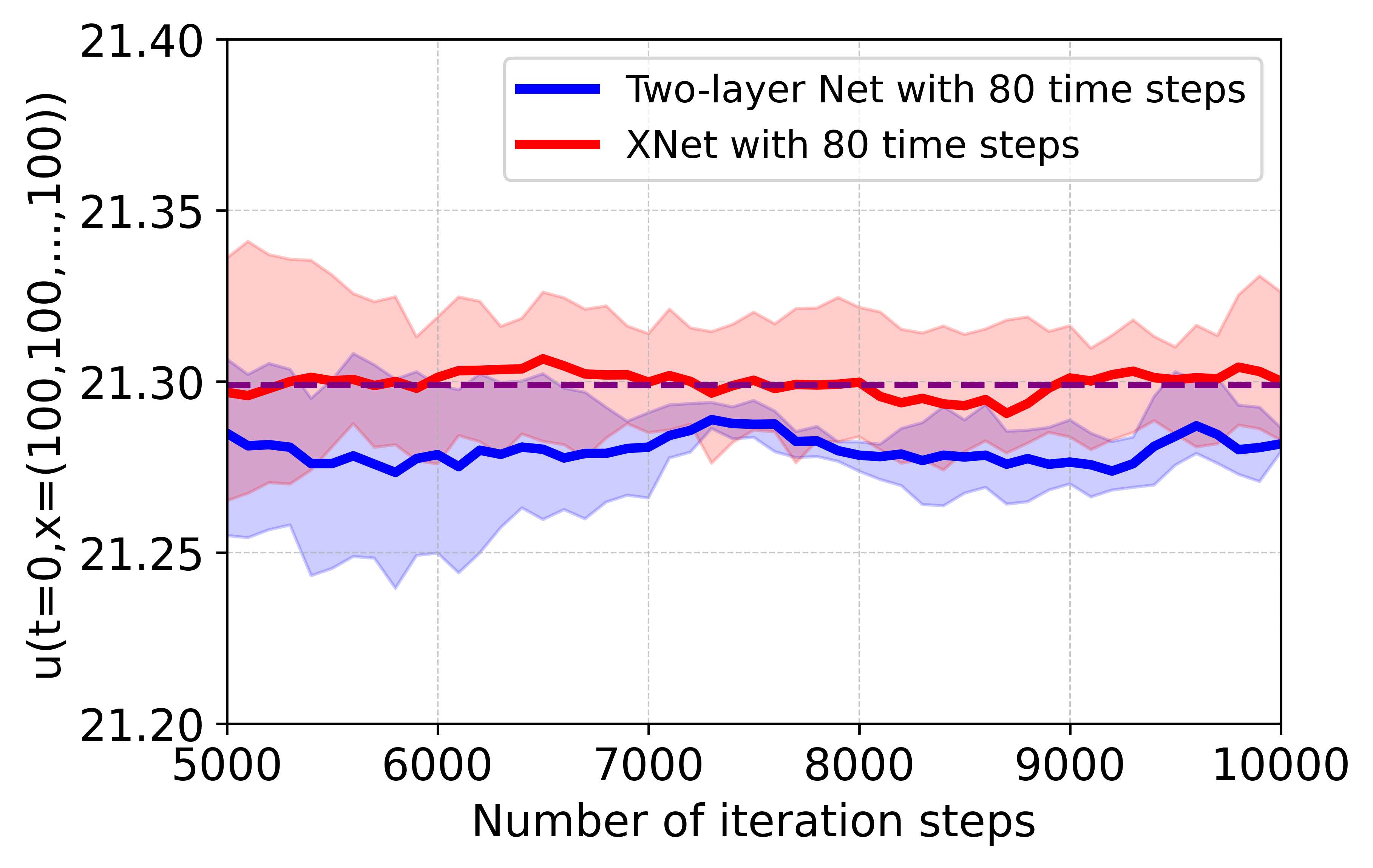}
		%\centerline{(b) Relative ${L^1}$-approximation error}
		\label{fig_dt_diffrate_80:second}
	\end{minipage}
	\vspace{-0.7cm}
	\centering
	\caption{Comparison of Two Network Architectures for Solving the PricingDiffrate Equation under 20-step-time Discretization and 80-step-time Discretization}
	\label{fig_dt_diffrate2}
\end{figure*}

From equation (\ref{diffrateth}), it is evident that the error introduced by temporal discretization is minimal, and the approximation error is almost exclusively related to the network's approximation capability and optimization errors.
\begin{equation}\label{diffrateth}
	\begin{aligned}
		\mathbb{E}\left|\frac{{u(0, \xi)-\theta_{{u_0}}}}{{u(0,\xi)}}  \right|^2\
		\leq &C\left\{ \frac{h}{21.299^2} +\inf _{\mu_0^\pi \in \mathcal{N}_0}  \mathbb{E}\left| Z_0-\theta_{\nabla{u_0}} \right|^2  \frac{h}{21.299^2} \right.\\
		&\left. +  \inf _{\phi_n^\pi \in \mathcal{N}_i} \sum_{i=0}^{N-1} \mathbb{E}\left|\mathbb{E}\left[\tilde{Z}_{t_n} \mid X_{t_n}^\pi, Y_{t_n}^\pi\right]-\phi_n\left(X_{t_n}^\pi, Y_{t_n}^\pi\right)\right|^2 \frac{h}{21.299^2}\right\}.
	\end{aligned}
\end{equation}
Consequently, as shown in Table \ref{dt_diffrate_results}, increasing the number of temporal steps does not yield significant improvement in computational accuracy for either architecture.
As shown in Table \ref{dt_diffrate_results} and Figure \ref{fig_dt_diffrate2}, for the time-step discretization $N=20,30,40,80$, implementing the Deep BSDE method using the XNet rather than FNNs enhances computational speed and significantly improves accuracy. This improvement can be attributed to XNet providing smaller neural network approximation errors and optimization errors in the Deep BSDE method.

In fact, we have already achieved very good results. The reason we do not further refine the time discretization is due to concerns that a significant increase in network parameters would introduce non-negligible optimization errors.
In the following section, we propose a continuous-time network architecture to eliminate the requirement for configuring a network at each temporal step. This approach addresses the challenge of significant parameter growth due to finer temporal discretization, thereby reducing optimization errors.

\begin{table}[htbp]
	\centering
	\caption{Numerical Results for PricingDiffrate Equation}
	\resizebox{\textwidth}{!}{%
		\begin{tabular}{ccccccccc}
			\hline \hline
			& \multicolumn{4}{c}{XNet} & \multicolumn{4}{c}{Feedforward neural networks} \\
			\midrule
			Time steps & Runtime (s)  & Value function & Relative Error  & Std. Deviation  
			& Runtime (s) & value function & Relative Error  & Std. Deviation \\
			\midrule
			20 & 69 & 2.1306e+01 & 3.3219e-04  & 4.2208e-03 & 96 &2.1260e+01 & 1.8144e-03 & 2.4084e-03 \\
			30 & 116 & 2.1302e+01 & 1.5146e-04 & 5.5465e-03 & 198  & 2.1279e+01 & 9.4557e-04 & 4.0696e-03 \\
			40 & 176 & 2.1303e+01 & 1.7159e-04 & 5.1489e-03 & 257 &2.1278e+01& 9.8029e-04 & 4.8573e-03 \\
			80 & 476 & 2.1304e+01 & 2.2257e-04 & 3.9802e-03 & 729&2.1280e+01 & 8.9073e-04 & 3.7753e-03 \\
			\hline \hline
		\end{tabular}%
	}
	\label{dt_diffrate_results}
\end{table}

\section{Continuous time models}\label{s5}
In this paper, the computational error in deep learning algorithms is categorized into four primary components: the approximation error induced by temporal time discretization, the approximation error arising from neural network representation, the generalization error governed by the number of training samples, and the optimization error associated with the number of network parameters. When the computational error is dominated by the approximation error induced by temporal discretization, increasing the temporal steps is necessary to achieve higher computational accuracy. However, in discrete-time network architectures, this inevitably increases the number of network parameters, reducing computational efficiency and increasing optimization error.
To address this issue, in this section, we apply the Deep BSDE method using both XNet and feedforward neural networks (FNNs) within continuous-time network architectures. In these continuous-time network architectures, the input layer includes an additional temporal dimension, resulting in a $d+1$-dimensional input. The output is the gradient function at each time step, which is $d$-dimensional. The difference lies in the fact that XNet has only one hidden layer ($d$-dimensional), while the FNNs comprise two hidden layers (each with $d+10$ dimensions).

\subsection{Allen-Cahn Equation}
In the previous section, when solving the Allen-Cahn equation (\ref{AllenCahneq}) using discrete-time network architectures, it was observed that the accuracy of the algorithm improved with finer temporal discretization (Table \ref{dt_allen_cahn_results}). However, the increase in network parameters resulted in higher optimization errors. Here, continuous-time network architectures are adopted. First, XNet possesses sufficient approximation capability, suggesting that the approximation error arising from the neural network is minimal. Second, since XNet has relatively few parameters, we assume that the optimization error is also minimal. Third, we sampled $640,000$ independent trajectories, indicating that the generalization error is small.
Under the assumption that the error in the Deep BSDE method is primarily dominated by the approximation error induced by temporal discretization, we can analyze the convergence rate of the Deep BSDE method with XNet.

\begin{table}[ht]
	\centering
	\caption{Numerical Results for Allen-Cahn Equation}
	\resizebox{\textwidth}{!}{%
		\begin{tabular}{ccccccccccc}
			\hline \hline
			& \multicolumn{5}{c}{XNet} & \multicolumn{5}{c}{Feedforward neural networks} \\
			\midrule
			Time steps & Runtime (s)&Value function & Relative Error & Error order  & Std. Deviation  
			& Runtime (s) &Value function & Relative Error & Error order & Std. Deviation \\
			\midrule
			10 & 47 & 5.3020e-02 & 4.1212e-03 & 
			 & 4.8446e-05 & 42 & 5.3020e-02 & 4.1269e-03 &
			  & 4.5878e-05 \\
			20 & 121 & 5.2906e-02 & 1.9607e-03 & 1.07 & 4.1509e-05 & 138 & 5.2907e-02 & 1.9917e-03 &  1.05 & 9.0286e-05 \\
			40 & 261 & 5.2842e-02 & 7.6389e-04 &1.36
			 & 4.2562e-05 & 367 & 5.2833e-02& 5.9284e-04 &1.75 & 7.4642e-05 \\
			80 & 843 & 5.2810e-02 & 1.5496e-04 & 2.30
			 & 3.3253e-05 & 1452&5.2828e-02 & 4.9686e-04 & 0.26 & 7.4642e-05 \\
			160 & 2004 & 5.2805e-02 & 4.8824e-05 & 1.67
			 & 5.1628e-05 & 3728& 5.2815e-02 & 2.4993e-04 &  0.99
			 & 1.4407e-04 \\
			\hline \hline
		\end{tabular}%
	}
	\label{ct_allen_cahn_results}
\end{table}

\begin{figure*}
	\centering
	\begin{minipage}[t]{0.478\linewidth}
		\centering
		\includegraphics[width=\textwidth]{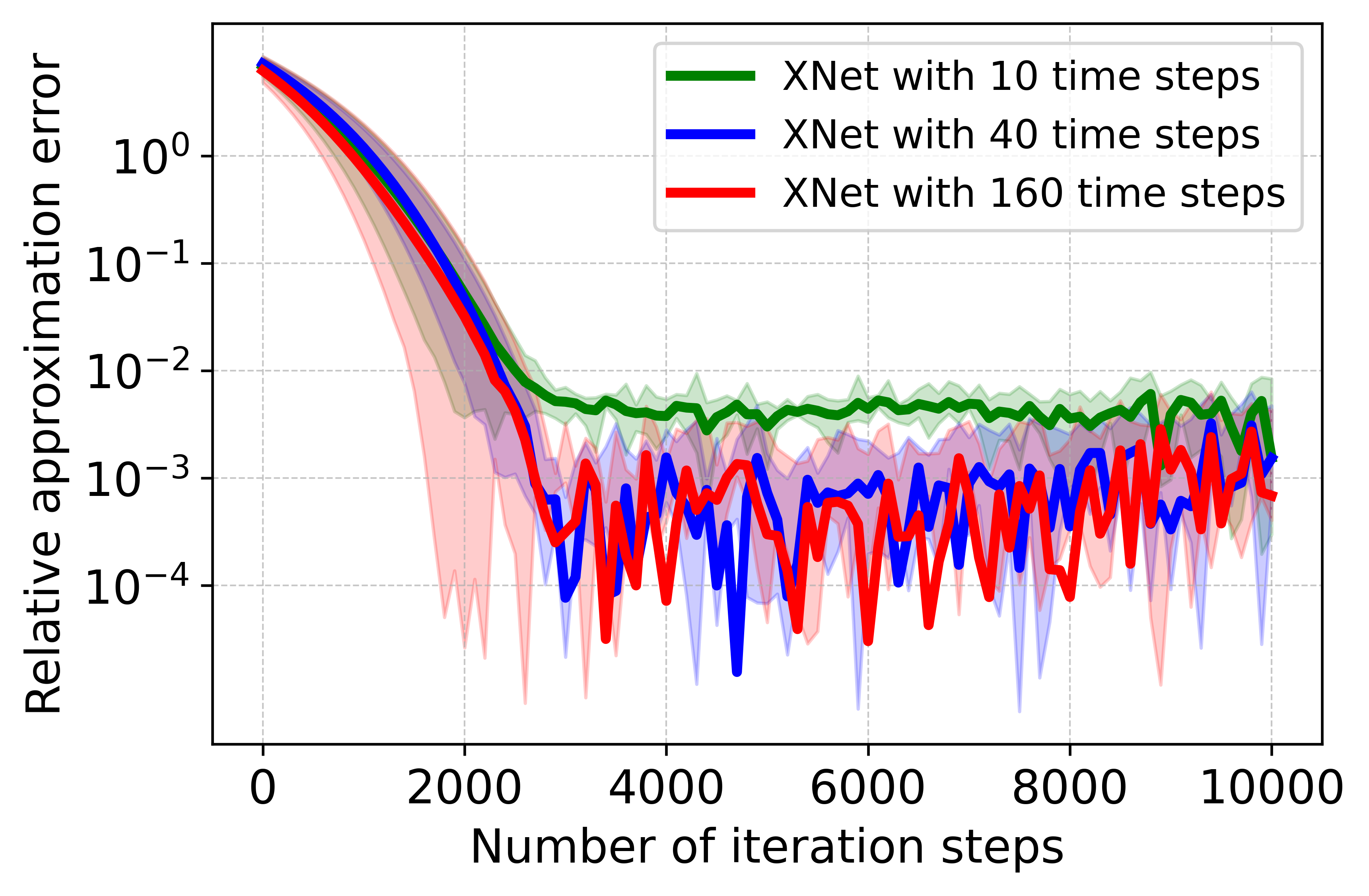}
		%\centerline{(a) Relative ${L^1}$-approximation error}
		\label{fig_ct_allencahn_10:first}
	\end{minipage}%
	\hfill
	\begin{minipage}[t]{0.5\linewidth}
		\centering
		\includegraphics[width=\textwidth]{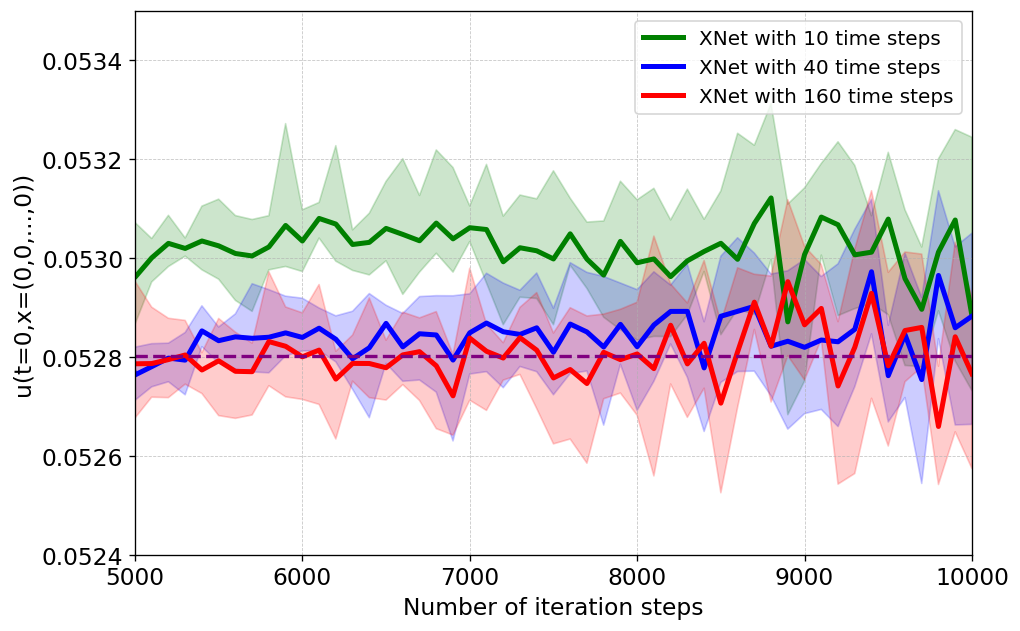}
		%\centerline{(b) Relative ${L^1}$-approximation error}
		\label{fig_ct_allencahn_160:second}
	\end{minipage}
	\vspace{-0.7cm}
	\centering
	\caption{Results of solving the Allen-Cahn Equation using the Deep BSDE method by XNet under $N$-time-step discretization, with $N=10$, $40$, and $160$.}
	\label{fig_ct_allencahn}
\end{figure*}

Table \ref{ct_allen_cahn_results} and Figure \ref{fig_ct_allencahn} demonstrate that the convergence order of the Deep BSDE method implemented using XNet is slightly less than second order. Notably, when the number of temporal steps reaches 40, FNNs no longer exhibit a discernible convergence order, whereas XNet maintains consistent convergence behavior. This can be attributed to the differential approximation capabilities: XNet possesses superior approximation capabilities, resulting in minimal neural network approximation errors and optimization errors, whereas FNNs lack this property.

With finer temporal discretization, the accuracy improved by both network implementations. The XNet achieves higher accuracy within shorter computational time and demonstrates greater robustness.

\subsection{Pricing of European financial derivatives with different interest rates for borrowing and lending (PricingDiffrate) equation}

\begin{table}[htbp]
	\centering
	\caption{Numerical Results for solving PricingDiffrate Equation}
	\resizebox{\textwidth}{!}{%
		\begin{tabular}{cccccccccccc}
			\hline \hline
			& \multicolumn{5}{c}{XNet} & \multicolumn{5}{c}{Feedforward neural networks} \\
			\midrule
			Time steps & Runtime (s)  & Value function & Relative Error& Error order  & Std. Deviation  
			& Runtime (s) & Value function & Relative Error &Error order & Std. Deviation \\
			\midrule
			10 & 55 & 2.1305e+01 & 2.6617e-04 & & 2.7172e-03 & 96 &2.1219e+01 & 3.7457e-03 & & 3.2896e-03 \\
			20 & 70 & 2.1296e+01 & 1.5515e-04 & 0.78& 2.5540e-03 & 142  & 2.1228e+01 & 3.3429e-03 & 0.16& 2.9154e-03 \\
			40 & 151 & 2.1302e+01 & 1.2577e-04 & 0.30&3.5567e-03 & 273 &2.1238e+01 & 2.8748e-03 &0.22 & 2.5155e-03 \\
			80 & 506 &  2.1301e+01 & 9.7900e-05 &0.36 & 1.4253e-03 & 729&2.1224e+01 & 3.5309e-03 & -0.30& 2.0813e-03 \\
			160 & 1558 & 2.1297e+01 & 9.0625e-05 &0.11 & 2.3293e-03 & 2736&2.1215e+01 & 3.9331e-03 & -0.16& 2.0182e-03 \\
			\hline \hline
		\end{tabular}%
	}
	\label{ct_diffrate_results}
\end{table}

\begin{figure}[H]
	\centering
	\begin{minipage}[t]{0.5\linewidth}
		\centering
		\includegraphics[width=\textwidth]{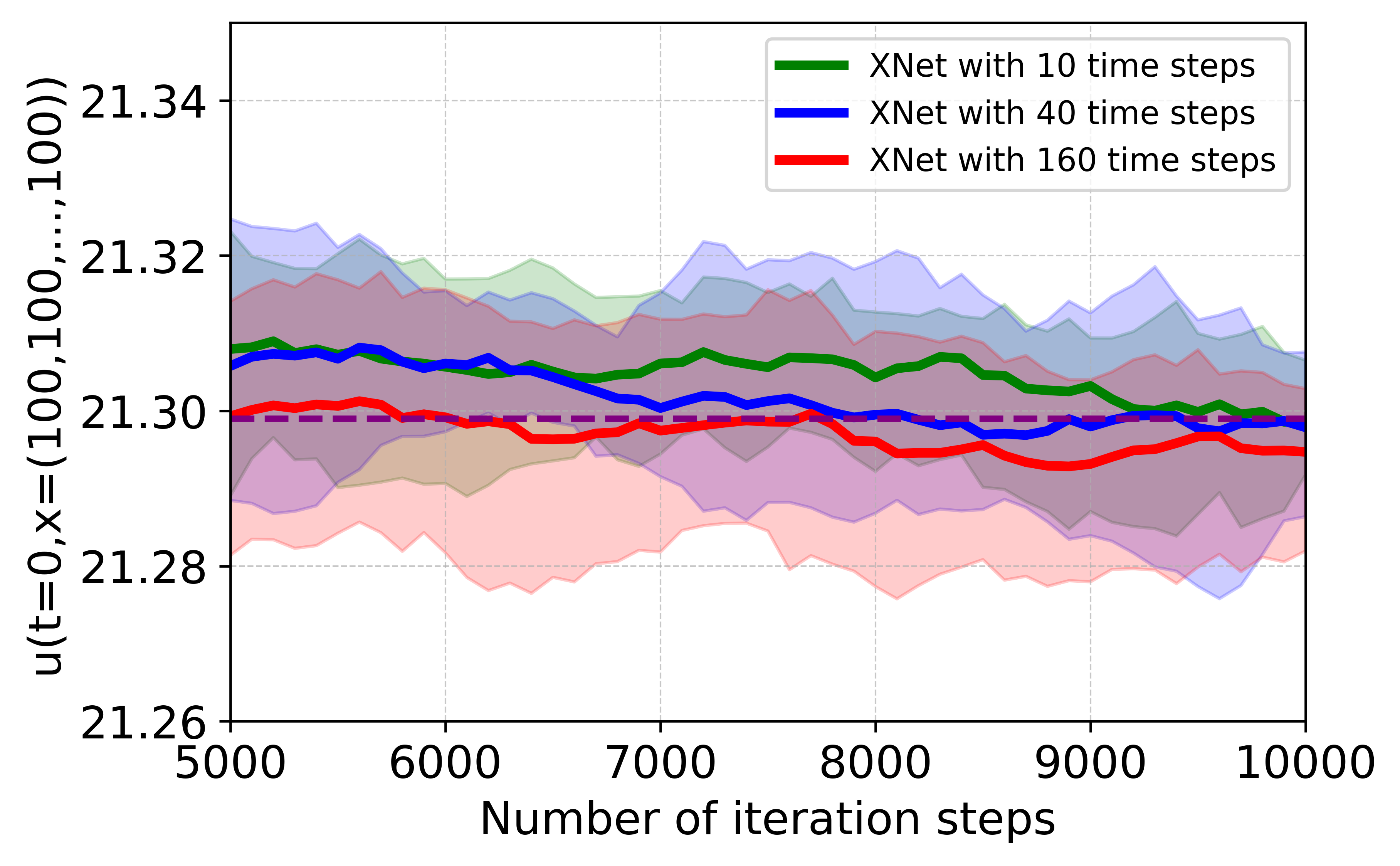}
		%\centerline{(a) Relative ${L^1}$-approximation error}
		\label{fig_ct_diffrate_20:first}
	\end{minipage}%
	\hfill
	\begin{minipage}[t]{0.5\linewidth}
		\centering
		\includegraphics[width=\textwidth]{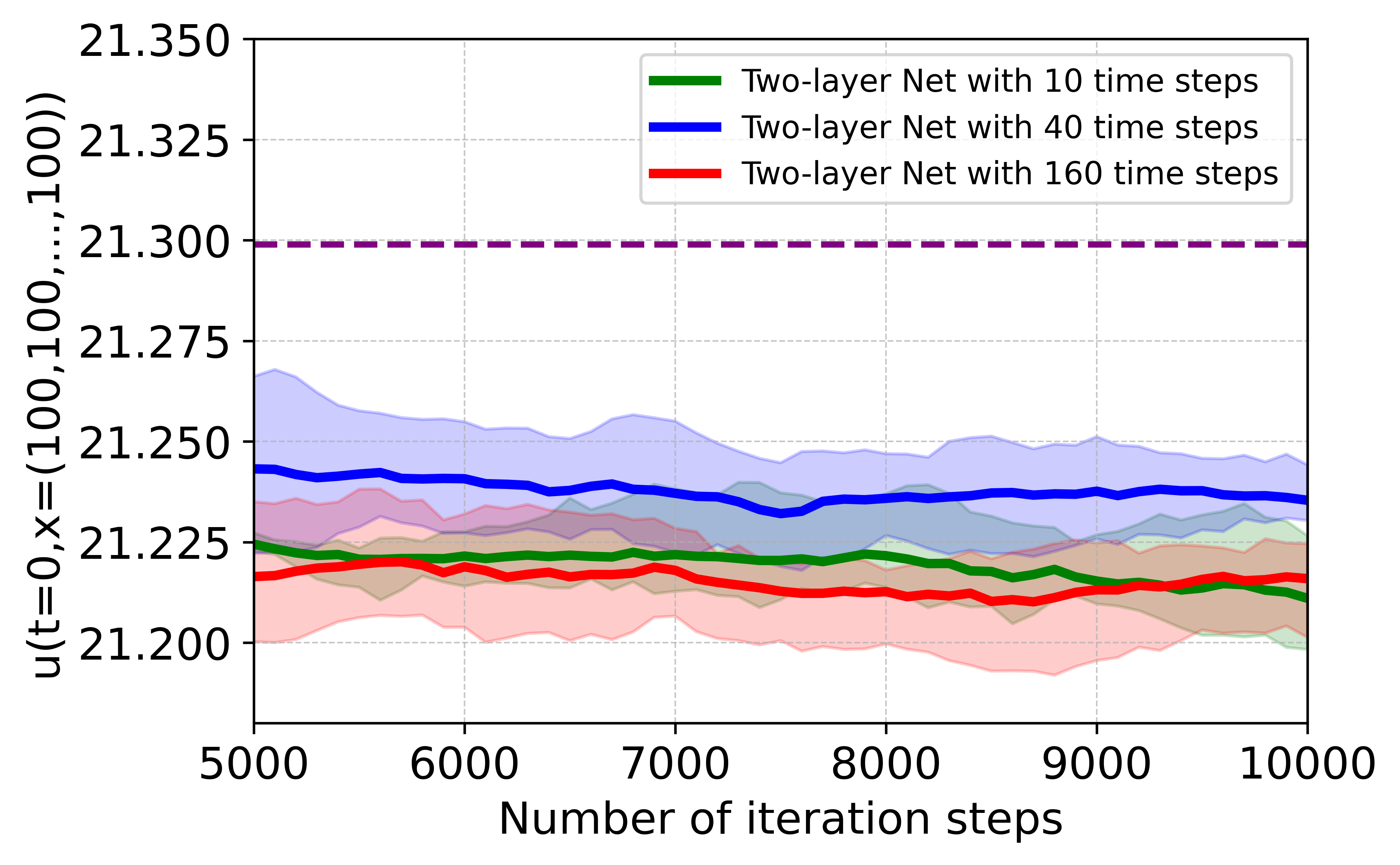}
		%\centerline{(b) Relative ${L^1}$-approximation error}
		\label{fig_ct_diffrate_20:second}
	\end{minipage}
	\vspace{-0.7cm}
	\centering
	\caption{Comparison of Two Network Architectures for Solving the PricingDiffrate under 10-time-step and 160-time-step Discretization}
	\label{fig_ct_diffrate1_1}
\end{figure}

The Deep BSDE method is also applied to solve the PricingDiffrate equation (\ref{EPDFeq}) with continuous-time implementations of the XNet and the FNNs.
Table \ref{ct_diffrate_results} and Figure \ref{fig_ct_diffrate1_1} demonstrate that, regardless of the temporal discretization used ($N=10, 20, 40, 80, 160$), the Deep BSDE method implemented with XNet consistently outperforms in terms of both computational speed and accuracy. These results align with the findings from the discrete-time models discussed in Section \ref{s4}.

Although introducing XNet allows the Deep BSDE method to achieve an accuracy approaching $9 \times 10^{-5}$, no clear error order is observed in this example. We speculate that the approximation error induced by temporal discretization is minimal, and that the computational error is likely dominated by the neural network's approximation capability or training error. To analyze the convergence rate, it is crucial to ensure that the  approximation errors, optimization errors, and training errors are all minimized.
To this end, we increase the batch size to reduce the generalization errors, and increase the number of basis functions in XNet to enhance the neural network's approximation capability.

\begin{table}[htbp]
	\centering
	\caption{Numerical Results for solving PricingDiffrate Equation by XNet}
	\resizebox{0.9\textwidth}{!}{%
		\begin{tabular}{cccccccc}
			\hline \hline
			Steps & Batch size & Basis Functions & Runtime (s)  & value function & Relative Error &Error order & Std. Deviation \\
			\midrule
			20 &64 & 100 & 70 & 2.1296e+01 & 1.5515e-04 & & 2.5540e-03 \\
			20 &256 & 100 & 98 & 2.1304e+01 & 2.4130e-04 & & 4.4412e-03 \\
			10 & 64 & 200 & 44 & 2.1295e+01 & 2.0725e-04 & & 3.6821e-03  \\
			20 & 64 & 200 & 98 & 2.1301e+01 & 8.9936e-05 & 1.20& 3.0920e-03 \\
			40 & 64 & 200 & 207 & 2.1298e+01 & 6.4311e-05 &0.48 & 1.2050e-03 \\
			80 & 64 & 200 & 670 &2.1300e+01 & 3.2679e-05 &0.98 & 2.5948e-03\\
			%160 & 64 & 200 & 1755 & 2.1300e+01 & 3.2442e-05 & &1.4426e-03 \\
			\hline \hline
		\end{tabular}%
	}
	\label{ct_diffrate_results_2}
\end{table}

\begin{figure}[H]
	\centering
	\begin{minipage}[t]{0.5\linewidth}
		\centering
		\includegraphics[width=\textwidth]{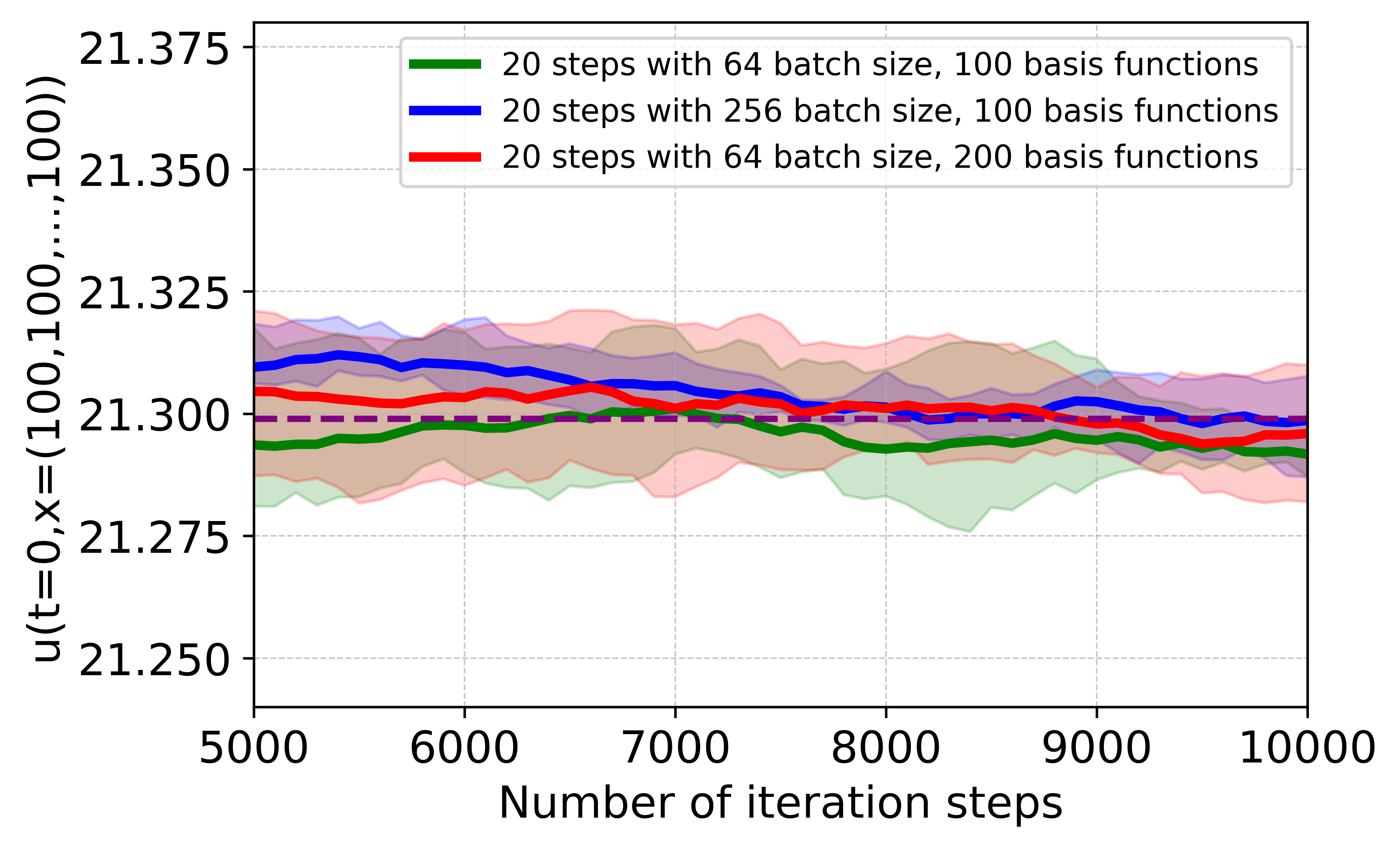}
		%\centerline{(a) Relative ${L^1}$-approximation error}
		\label{fig_ct_diffrate_20:first}
	\end{minipage}%
	\hfill
	\begin{minipage}[t]{0.5\linewidth}
		\centering
		\includegraphics[width=\textwidth]{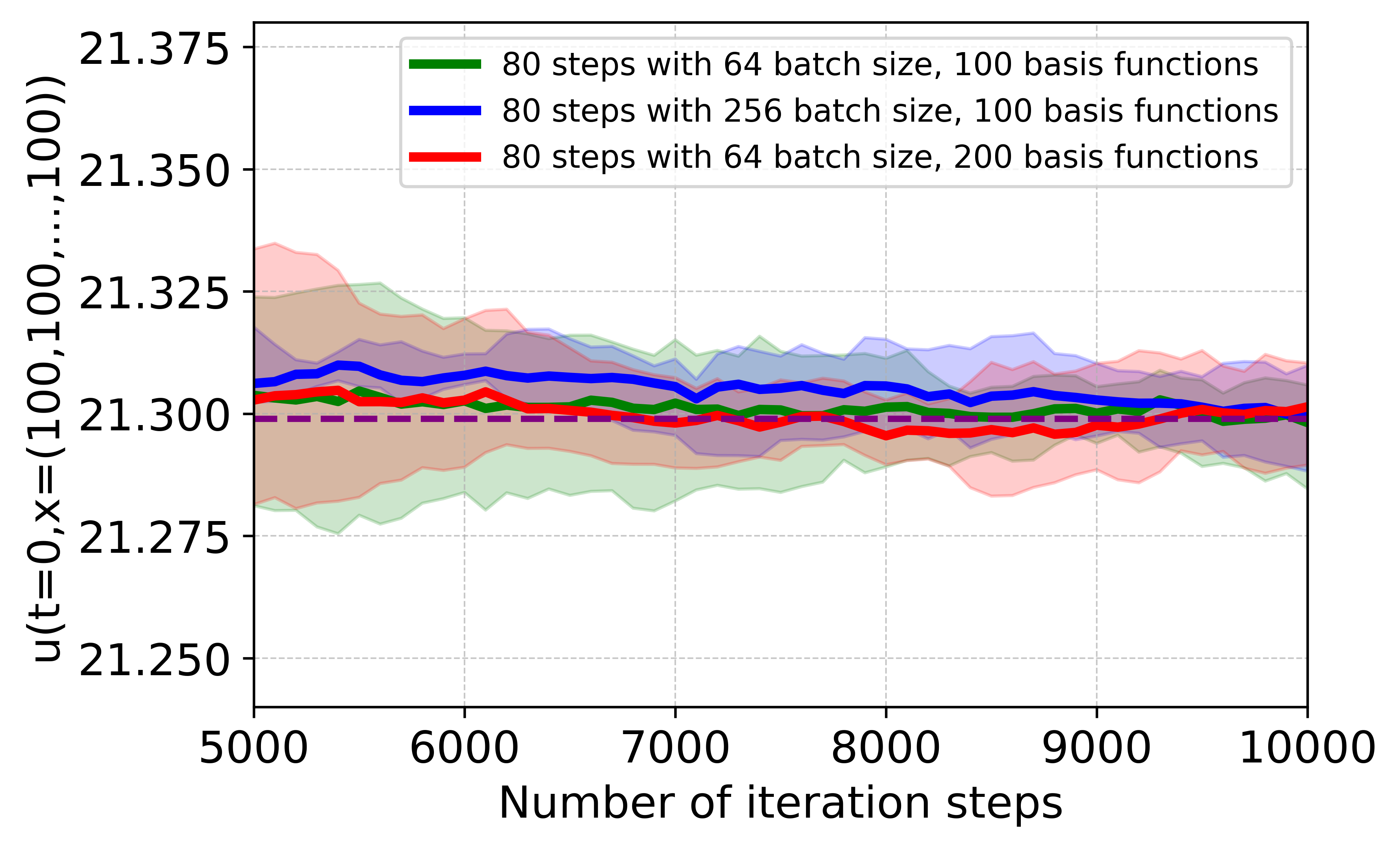}
		%\centerline{(b) Relative ${L^1}$-approximation error}
		\label{fig_ct_diffrate_20:second}
	\end{minipage}
	\vspace{-0.7cm}
	\centering
	\caption{Solving the PricingDiffrate Equation by XNet under various settings with 20-step-time Discretization and 80-step-time Discretization.}
	\label{fig_ct_diffrate2}
\end{figure}

As shown in Table \ref{ct_diffrate_results_2} and Figure \ref{fig_ct_diffrate2}, with a time-step discretization of $20$ or $80$, we observe the following:
On one hand, when the batch size reaches 256, the generalization error arising from the training samples no longer significantly contributes to the overall computational error.
On the other hand, by increasing the number of basis functions in XNet, the accuracy improves further.

As shown in Table \ref{ct_diffrate_results_2}, by increasing the number of basis functions in XNet to 200, we can clearly observe that the computational accuracy improves consistently as the time-step discretization increases. When the temporal discretization reaches 80, the relative error of the Deep BSDE method with XNet achieves $3.2679 \times 10^{-5}$, which significantly exceeds the accuracy achieved with feedforward neural networks ($3.9331 \times 10^{-3}$).

\section{Conclusion}
This work establishes a comprehensive theoretical and computational framework for Deep BSDE methods applied to high-dimensional semilinear parabolic PDEs with non-Lipschitz generators. We have rigorously extended convergence theory beyond the classical Lipschitz framework to encompass Allen-Cahn equations with cubic nonlinearity and Hamilton-Jacobi-Bellman equations with quadratic gradient growth, providing the theoretical justification for the empirical success observed in these applications. The introduction of XNet architecture represents a significant computational advancement, achieving $\mathcal{O}(L)$ parameter complexity while maintaining superior approximation capabilities compared to traditional $\mathcal{O}(HL^2)$ feedforward networks. In discrete-time implementations, XNet demonstrates substantial improvements in both computational efficiency and solution accuracy across all tested configurations, with faster convergence and reduced relative errors for both Allen-Cahn and financial derivative pricing problems. The continuous-time framework reveals even more pronounced advantages: XNet significantly reduces neural network approximation errors and optimization errors compared to feedforward networks, enabling clearer observation of convergence behavior with rates approaching 1.6 for Allen-Cahn equations, while feedforward networks fail to maintain consistent convergence beyond 40 time steps. For the financial derivative pricing problem, XNet achieves relative errors of $3.27 \times 10^{-5}$, demonstrating markedly superior convergence orders compared to feedforward implementations. The theoretical convergence guarantees, combined with the demonstrated computational advantages of XNet in both discrete and continuous-time settings, establish a robust foundation for tackling high-dimensional PDE problems in scientific computing, with immediate applications in mathematical finance, materials science, and optimal control theory.

\clearpage
%\phantomsection 
%\addcontentsline{toc}{section}{References}
%\begin{thebibliography}{99}
%\end{thebibliography}

\bibliography{refbib}

@book{weishu1,
  title={A proof that artificial neural networks overcome the curse of dimensionality in the numerical approximation of Black--Scholes partial differential equations},
  author={Grohs, Philipp and Hornung, Fabian and Jentzen, Arnulf and Von Wurstemberger, Philippe},
  volume={284},
  number={1410},
  year={2023},
  publisher={American Mathematical Society}
}

@article{PINN,
  title={Physics-informed neural networks: A deep learning framework for solving forward and inverse problems involving nonlinear partial differential equations},
  author={Raissi, Maziar and Perdikaris, Paris and Karniadakis, George E},
  journal={Journal of Computational physics},
  volume={378},
  pages={686--707},
  year={2019},
  publisher={Elsevier}
}

@article{Shin,
  title={On the convergence of physics informed neural networks for linear second-order elliptic and parabolic type PDEs},
  author={Shin, Yeonjong and Darbon, Jerome and Karniadakis, George Em},
  journal={arXiv preprint arXiv:2004.01806},
  year={2020}
}

@article{Mishra,
  title={Estimates on the generalization error of physics-informed neural networks for approximating a class of inverse problems for PDEs},
  author={Mishra, Siddhartha and Molinaro, Roberto},
  journal={IMA Journal of Numerical Analysis},
  volume={42},
  number={2},
  pages={981--1022},
  year={2022},
  publisher={Oxford University Press}
}

@article{PINN_2024_1,
  title={AT-PINN: Advanced time-marching physics-informed neural network for structural vibration analysis},
  author={Chen, Zhaolin and Lai, Siu-Kai and Yang, Zhichun},
  journal={Thin-Walled Structures},
  volume={196},
  pages={111423},
  year={2024},
  publisher={Elsevier}
}

@article{PINN_2024_2,
  title={NAS-PINN: neural architecture search-guided physics-informed neural network for solving PDEs},
  author={Wang, Yifan and Zhong, Linlin},
  journal={Journal of Computational Physics},
  volume={496},
  pages={112603},
  year={2024},
  publisher={Elsevier}
}

@article{DGM,
  title={DGM: A deep learning algorithm for solving partial differential equations},
  author={Sirignano, Justin and Spiliopoulos, Konstantinos},
  journal={Journal of computational physics},
  volume={375},
  pages={1339--1364},
  year={2018},
  publisher={Elsevier}
}

@article{DGM_2024,
  title={Convergence of the Deep Galerkin Method for Mean Field Control Problems},
  author={Hofgard, William and Sun, Jingruo and Cohen, Asaf},
  journal={arXiv preprint arXiv:2405.13346},
  year={2024}
}

@article{Ritz,
  title={The deep Ritz method: a deep learning-based numerical algorithm for solving variational problems},
  author={Yu, Bing and others},
  journal={Communications in Mathematics and Statistics},
  volume={6},
  number={1},
  pages={1--12},
  year={2018},
  publisher={Springer}
}

@article{Ritz_2024,
  title={Deep Ritz Method for Elliptical Multiple Eigenvalue Problems},
  author={Ji, Xia and Jiao, Yuling and Lu, Xiliang and Song, Pengcheng and Wang, Fengru},
  journal={Journal of Scientific Computing},
  volume={98},
  number={2},
  pages={48},
  year={2024},
  publisher={Springer}
}

@article{XNet1,
  title={Cauchy activation function and xnet},
  author={Li, Xin and Xia, Zhihong and Zhang, Hongkun},
  journal={arXiv preprint arXiv:2409.19221},
  year={2024}
}

@article{XNet2,
  title={Enhancing Neural Function Approximation: The XNet Outperforming KAN},
  author={Li, Xin and Zheng, Xiaotao and Xia, Zhihong},
  journal={arXiv preprint arXiv:2501.18959},
  year={2025}
}

@article{E2017,
  title={Deep learning-based numerical methods for high-dimensional parabolic partial differential equations and backward stochastic differential equations},
  author={Han, Jiequn and Jentzen, Arnulf and others},
  journal={Communications in mathematics and statistics},
  volume={5},
  number={4},
  pages={349--380},
  year={2017},
  publisher={Springer}
}

@article{han2018solving,
  title={Solving high-dimensional partial differential equations using deep learning},
  author={Han, Jiequn and Jentzen, Arnulf and E, Weinan},
  journal={Proceedings of the National Academy of Sciences},
  volume={115},
  number={34},
  pages={8505--8510},
  year={2018},
  publisher={National Academy of Sciences}
}

@inproceedings{davey2024deep,
  title={Deep Neural Network Solver for HJB Equations},
  author={Davey, Ashley and Zheng, Harry},
  booktitle={APCA International Conference on Automatic Control and Soft Computing},
  pages={488--502},
  year={2024},
  organization={Springer}
}

@article{Beck2,
  title={Deep splitting method for parabolic PDEs},
  author={Beck, Christian and Becker, Sebastian and Cheridito, Patrick and Jentzen, Arnulf and Neufeld, Ariel},
  journal={SIAM Journal on Scientific Computing},
  volume={43},
  number={5},
  pages={A3135--A3154},
  year={2021},
  publisher={SIAM}
}

@article{DBDP,
  title={Deep backward schemes for high-dimensional nonlinear PDEs},
  author={Hur{\'e}, C{\^o}me and Pham, Huy{\^e}n and Warin, Xavier},
  journal={Mathematics of Computation},
  volume={89},
  number={324},
  pages={1547--1579},
  year={2020}
}

@inproceedings{Peng1,
  title={Backward stochastic differential equations and quasilinear parabolic partial differential equations},
  author={Pardoux, Etienne and Peng, Shige},
  booktitle={Stochastic Partial Differential Equations and Their Applications: Proceedings of IFIP WG 7/1 International Conference University of North Carolina at Charlotte, NC June 6--8, 1991},
  pages={200--217},
  year={2005},
  organization={Springer}
}

@article{Peng2,
  title={Forward-backward stochastic differential equations and quasilinear parabolic PDEs},
  author={Pardoux, Etienne and Tang, Shanjian},
  journal={Probability theory and related fields},
  volume={114},
  pages={123--150},
  year={1999},
  publisher={Springer}
}

@article{Convergence-of-the-deep-BSDE-method,
  title={Convergence of the deep BSDE method for coupled FBSDEs},
  author={Han, Jiequn and Long, Jihao},
  journal={Probability, Uncertainty and Quantitative Risk},
  volume={5},
  number={1},
  pages={5},
  year={2020},
  publisher={Springer}
}

@article{weishu,
  title={Multilayer feedforward networks are universal approximators},
  author={Hornik, Kurt and Stinchcombe, Maxwell and White, Halbert},
  journal={Neural networks},
  volume={2},
  number={5},
  pages={359--366},
  year={1989},
  publisher={Elsevier}
}

@incollection{zhang2017backward,
  title={Backward stochastic differential equations},
  author={Zhang, Jianfen},
  booktitle={Backward Stochastic Differential Equations: From Linear to Fully Nonlinear Theory},
  pages={79--99},
  year={2017},
  publisher={Springer}
}

@article{Xiao,
  title={Deep learning based on randomized quasi-Monte Carlo method for solving linear Kolmogorov partial differential equation},
  author={Xiao, Jichang and Fu, Fengjiang and Wang, Xiaoqun},
  journal={Journal of Computational and Applied Mathematics},
  pages={116088},
  year={2024},
  publisher={Elsevier}
}

@article{Importance_sampling,
  title={Importance sampling: a review},
  author={Tokdar, Surya T and Kass, Robert E},
  journal={Wiley Interdisciplinary Reviews: Computational Statistics},
  volume={2},
  number={1},
  pages={54--60},
  year={2010},
  publisher={Wiley Online Library}
}

@article{QMC,
  title={Quasi-monte carlo methods},
  author={Sobo{\'l}, IM},
  journal={Progress in Nuclear Energy},
  volume={24},
  number={1-3},
  pages={55--61},
  year={1990},
  publisher={Elsevier}
}

@article{Gibbs,
  title={Gibbs sampling},
  author={Gelfand, Alan E},
  journal={Journal of the American statistical Association},
  volume={95},
  number={452},
  pages={1300--1304},
  year={2000},
  publisher={Taylor \& Francis}
}

@misc{Parameter_optimization,
  title={Parameter optimization in neural networks},
  author={Katanforoosh, Kunin and Kunin, D and Ma, J},
  year={2019},
  publisher={Retrieved from deeplearning. ai: https://www. deeplearning. ai/ai-notes~…}
}

@article{Shen_Z,
  title={Neural network approximation: Three hidden layers are enough},
  author={Shen, Zuowei and Yang, Haizhao and Zhang, Shijun},
  journal={Neural Networks},
  volume={141},
  pages={160--173},
  year={2021},
  publisher={Elsevier}
}

@article{FENZHI1,
  title={Counterparty risk valuation: A marked branching diffusion approach},
  author={Henry-Labordere, Pierre},
  journal={arXiv preprint arXiv:1203.2369},
  year={2012}
}

@article{FENZHI2,
  title={Branching diffusion representation of semilinear PDEs and Monte Carlo approximation},
  author={Henry-Labord{\`e}re, Pierre and Oudjane, Nadia and Tan, Xiaolu and Touzi, Nizar and Warin, Xavier},
  journal={55 1 ANNALES DE L’INSTITUT HENRI POINCAR{\'E} PROBABILIT{\'E}S ET STATISTIQUES Vol. 55, No. 1 (February, 2019) 1--607},
  volume={55},
  number={1},
  pages={184--210},
  year={2019}
}

@article{FENZHI3,
  title={A numerical algorithm for a class of BSDEs via the branching process},
  author={Henry-Labordere, Pierre and Tan, Xiaolu and Touzi, Nizar},
  journal={Stochastic Processes and their Applications},
  volume={124},
  number={2},
  pages={1112--1140},
  year={2014},
  publisher={Elsevier}
}

@article{Bergman4,
  title={Option pricing with differential interest rates},
  author={Bergman, Yaacov Z},
  journal={The Review of Financial Studies},
  volume={8},
  number={2},
  pages={475--500},
  year={1995},
  publisher={Oxford University Press}
}

@article{zhang2004numerical,
  title={A numerical scheme for BSDEs},
  author={Zhang, Jianfeng},
  journal={The annals of applied probability},
  volume={14},
  number={1},
  pages={459--488},
  year={2004},
  publisher={Institute of Mathematical Statistics}
}

@article{imkeller2010path,
  title={Path regularity and explicit convergence rate for BSDE with truncated quadratic growth},
  author={Imkeller, Peter and Dos Reis, Gon{\c{c}}alo},
  journal={Stochastic Processes and their Applications},
  volume={120},
  number={3},
  pages={348--379},
  year={2010},
  publisher={Elsevier}
}

@article{richou2012markovian,
  title={Markovian quadratic and superquadratic BSDEs with an unbounded terminal condition},
  author={Richou, Adrien},
  journal={Stochastic Processes and their Applications},
  volume={122},
  number={9},
  pages={3173--3208},
  year={2012},
  publisher={Elsevier}
}

@article{cai2024soc,
  title={Soc-martnet: A martingale neural network for the hamilton-jacobi-bellman equation without explicit inf H in stochastic optimal controls},
  author={Cai, Wei and Fang, Shuixin and Zhou, Tao},
  journal={arXiv preprint arXiv:2405.03169},
  year={2024}
}

@article{cai2024martingale,
  title={Martingale deep learning for very high dimensional quasi-linear partial differential equations and stochastic optimal controls},
  author={Cai, Wei and Fang, Shuixin and Zhang, Wenzhong and Zhou, Tao},
  journal={arXiv preprint arXiv:2408.14395},
  year={2024}
}

@article{bouchard2004discrete,
  title={Discrete-time approximation and Monte-Carlo simulation of backward stochastic differential equations},
  author={Bouchard, Bruno and Touzi, Nizar},
  journal={Stochastic Processes and their applications},
  volume={111},
  number={2},
  pages={175--206},
  year={2004},
  publisher={Elsevier}
}

@article{Zhang2008time,
  	title={TIME DISCRETIZATION AND MARKOVIAN ITERATION FOR COUPLED FBSDES},
  author={BENDER, CHRISTIAN and ZHANG, JIANFENG},
 journal={The Annals of Applied Probability},
  volume={18},
  number={1},
  pages={143--177},
  year={2008}
}

@article{chassagneux2016numerical,
  title={NUMERICAL SIMULATION OF QUADRATIC BSDES},
  author={Chassagneux, Jean-Fran{\c{c}}ois and Richou, Adrien},
  journal={The Annals of Applied Probability},
  pages={262--304},
  year={2016},
  publisher={JSTOR}
}

@article{kloeden1995numerical,
  title={Numerical solution of stochastic differential equations (Peter E. Kloeden and Eckhard Platen)},
  author={Gelbrich, Matthias and R{\"o}misch, Werner},
  journal={SIAM Review},
  volume={37},
  number={2},
  pages={272--275},
  year={1995},
  publisher={SIAM}
}
\bibliographystyle{siam} 

\newpage
\appendix
\section*{Appendix A: Assumptions for Deep BSDE Method Convergence Analysis}

This appendix provides the mathematical assumptions required for the convergence analysis of Deep BSDE methods under different generator conditions.
For Lipschitz generators, the convergence analysis in Section 3.1 requires Assumptions \ref{assumption_1} and \ref{assumption_3}.
For Allen-Cahn type equations, the convergence analysis in Section 3.2 uses Assumptions \ref{assumption_1} and \ref{assumption_3}, where the cubic nonlinearity is handled through boundedness properties established in Lemma \ref{lem:L2boundY}.
For HJB type equations, the convergence analysis in Section 3.3 requires Assumption \ref{assumption_1} (with H$_2$ replaced by the conditions in Assumption \ref{ass:quadratic}) and Assumption \ref{assumption_3}.

{
We adopt the notation $\Delta x = x_1 - x_2$, $\Delta y = y_1 - y_2$, and $\Delta z = z_1 - z_2$. The constants $\mu_0, \sigma_0$, and $g_0$ are non-negative real numbers representing the bounds at the origin or the intercept terms.}

\begin{assumption}\label{assumption_1}
\

H$_1.$ The coefficients $\mu, \sigma, g$ satisfy standard regularity and bounded conditions. There exist (possibly negative) constants $k_\mu$ such that
$$
\begin{aligned}
\left[\mu(t,x_1)-\mu(t,x_2)\right]^\mathrm{T}\Delta x&\leq k_\mu|\Delta x|^2.
\end{aligned}
$$
there are non-negative constants $K, \sigma_x$, $f_x$, $f_z$, and $g_x$ such that
$$
\begin{aligned}
|\mu(t,x_1)-\mu(t,x_2)|^2 &\leq K|\Delta x|^2, \quad |\mu(t,x)|^{2}\leq \mu_{0}+K|x|^{2},\\
|\sigma(t,x_{1})-\sigma(t,x_{2})|^{2}&\leq\sigma_{x}|\Delta x|^{2}, \quad |\sigma(t,x)|^{2}\leq\sigma_{0}+\sigma_{x}|x|^{2}\\
|g(x_1)-g(x_2)|^2&\leq g_x|\Delta x|^2, \quad |g(x)|^{2}\leq g_{0}+g_{x}|x|^{2}.
\end{aligned}
$$

H$_2.$ $f$ is uniformly lipschitz continuous with respect $to$ $( x, y, z).$ In particular,
There exist (possibly negative) constants $k_f$ such that
$$
\begin{aligned}
\left[f(t,x,y_1,z)-f(t,x,y_2,z)\right]\Delta y&\leq k_f|\Delta y|^2.
\end{aligned}
$$
There are non-negative constants $K, \mu_y, \sigma_x, \sigma_y$, $f_x$, $f_z$, and $g_x$ such that
$$
\begin{aligned}
|f(t,x_1,y_1,z_1)-f(t,x_2,y_2,z_2)|^2&\leq f_x|\Delta x|^2+K|\Delta y|^2+f_z|\Delta z|^2,\\
|f(t,x,y,z)|^{2}&\leq f_{0}+f_{x}|x|^{2}+K|y|^{2}+f_{z}|z|^{2}.
\end{aligned}
$$

H$_3.$ $\mu, \sigma, f$ are uniformly Hölder-$\frac12$ continuous with respect to t.

\end{assumption}

\begin{assumption}\label{assumption_3}
One of the following five cases holds: 

H$_1.$ Small time duration, that is, T is small.

H$_2.$ Weak coupling of $Y$ into the forward SDE (\ref{Xt}), that is, $\mu_y$ and $\sigma_{y}$ are small. In particular, if $\mu_{y}=\sigma_{y}=0,$ then the forward equation does not depend on the
backward one and, thus, Eqs. (\ref{Xt}) and (\ref{Yt}) are decoupled.

H$_3.$ Weak coupling of $X$ into the backward SDE (\ref{Yt}), that is, $f_x$ and $g_x$ are small. In particular, if $f_{x}= g_{x}= 0$, then the backward equation does not depend on the forward one and, thus, Eqs. (\ref{Xt}) and (\ref{Yt}) are also decoupled. In fact, in this case, $Z=0$ and (\ref{Yt}) reduces to an ODE.

H$_4.$ $f$ is strongly decreasing in $y$, that is, $k_{f}$ is very negative. 

H$_5.$ $\mu$ is strongly decreasing in $x$, that is, $k_{\mu}$ is very negative.
\end{assumption}

\begin{assumption}[Quadratic Gradient Growth Framework]\label{ass:quadratic}
\

\textbf{H$_1^*$.} For any $0 \leq t \leq T$, the functions $\mu(t, \cdot)$, $\sigma(t, \cdot)$ are differentiable and their derivatives are uniformly Lipschitz with Lipschitz constant $K$ independent of $t$. 
In other words, $\sigma \in B_m^{m \times d}$ and $\mu \in B_m^{m \times 1}$. 
There exists a positive constant $c$ such that
\begin{equation}\label{eq:hx0}
    y^{T} \sigma(t, x) \sigma^{T}(t, x) y \geq c|y|^{2}, 
    \quad x, y \in \mathbb{R}^{m}, \; t \in [0, T].
\end{equation}

\textbf{H$_2^*$.} For the generator $f(t,x,y,z)$ with quadratic growth, there exists $C_f \in R_+$ such that:
\begin{align}
|f(t,x,y_1,z_1) - f(t,x,y_2,z_2)| &\leq  C_f|y_1 - y_2| +C_f(1 + |z_1| + |z_2|)|z_1 - z_2|, \\
|f(t,x_1,y,z) - f(t,x_2,y,z)| &\leq  C_f(1 + |y| + |z|^2)|x_1 - x_2| , \\
|f(t,x,y,z)| &\leq C_f(1 + |y| + |z|^2).
\end{align}
Hypothesis above holds. f is differentiable in $(x, y,z)$, and
\begin{align}
\nabla_x |f(t,x,y,z) | &\leq  C_f (1+|y| + |z|^2), \\
\nabla_y |f(t,x,y,z) | &\leq  C_f, \\
\nabla_z |f(t,x,y,z) | &\leq  C_f(1+|z|).
\end{align}

\textbf{H$_3^*$.} Moment boundedness: There exists $p > 2$ such that
$$\sup_{0 \leq t \leq T} \mathbb{E}[|Y_t|^p + |Z_t|^p] < \infty.$$

\textbf{H$_4^*$.} Neural network projection property: The approximation $Z_n^{\theta,\pi} = \varphi_N^\theta(Z_n^\pi)$ satisfies uniform bounds and projection regularity.
\end{assumption}

\section*{Appendix B: Proof of Boundedness Properties for Double-Well Dynamics}

This appendix provides the detailed proof of Lemma \ref{lem:L2boundY}, which establishes the crucial boundedness properties for the double-well dynamics. The boundedness result is essential for proving convergence of the Deep BSDE method for Allen-Cahn equations in Section 3.2.

\begin{proof}
The BSDE \eqref{eq:BSDE-Allen-Cahn} can be written in differential form as
\[
dY_t = (Y_t^3 - Y_t)\,dt + Z_t\,dW_t .
\]
Applying It\^o's formula to $Y_t^2$, we obtain
\[
d(Y_t^2)
= 2Y_t\,dY_t + |Z_t|^2\,dt
= \bigl( -2(Y_t^2 - Y_t^4) + |Z_t|^2 \bigr)\,dt
  + 2Y_t Z_t\,dW_t .
\]
Integrating from $t$ to $T$ yields
\[
Y_T^2 - Y_t^2
= \int_t^T \bigl( -2(Y_s^2 - Y_s^4) + |Z_s|^2 \bigr)\,ds
  + \int_t^T 2Y_s Z_s\,dW_s .
\]
Taking expectations and using the martingale property of the stochastic
integral, we obtain
\[
\mathbb E[Y_t^2]
= \mathbb E[Y_T^2]
  + \int_t^T \mathbb E\bigl( 2(Y_s^2 - Y_s^4) - |Z_s|^2 \bigr)\,ds .
\]
Since for all $y\in\mathbb R$,
\[
y^2 - y^4 = y^2(1-y^2) \le \tfrac14 ,
\]
we deduce that
\[
\mathbb E[Y_t^2]
\le \mathbb E[Y_T^2] + \frac{T-t}{2}
\le \mathbb E[g(X_T)^2] + \frac{T}{2} .
\]
This proves the uniform $L^2$-boundedness of $(Y_t)_{t\in[0,T]}$.
The bound on $\mathbb E|Y_t|$ follows immediately from the
Cauchy--Schwarz inequality.
\end{proof}

\section*{Appendix C: Convergence for HJB-type equations}

This appendix provides the complete convergence analysis for the Deep BSDE method applied to HJB-type equations with quadratic gradient growth generators. The main result establishes convergence rates under appropriate regularity assumptions.

We begin by establishing fundamental error bounds for the discrete approximation schemes.

\begin{lemma}[Forward Process Error Bound]\label{lemma:Truncation_X_Convergence}
Assume that $H_1$ in Assumption \ref{assumption_1} holds. Then the SDE 
\[
dX_t = \mu(t,X_t)\,dt + \sigma(t,X_t)\,dW_t, \qquad X_0 = x \in \mathbb{R}^d,
\] 
admits a unique solution. Moreover, we have
$$\mathbb{E}|\delta X_n^\pi|^2 \leq Ch,$$
where {$\delta X_n^\pi = {X}_{t_n} - X_n^\pi$}, and $C$ is independent of $h$.
\end{lemma}

{For the proof of this Lemma \ref{lemma:Truncation_X_Convergence}, we refer the reader to \cite{chassagneux2016numerical, imkeller2010path, kloeden1995numerical}.}
The core technical challenge lies in analyzing the convergence of the Deep BSDE system \eqref{system} to the truncated reference system \eqref{eq:perturbed_BTZ}. Before proving Theorem \ref{Th:estimation_HJB} and \ref{th:DBSDE_conv_PBTZ_brief}, we first introduce a proposition and a lemma.

% \begin{proof}
% Using the standard analysis for Euler schemes with Lipschitz coefficients:
% \begin{equation}
% \begin{aligned}
% \mathbb{E}\big|\delta X_{n}^\pi\big|^2 
% &= \mathbb{E}\big|\delta X_{{n-1}}^\pi\big|^2 
%   + \mathbb{E} \left[ \int_{t_{n-1}}^{t_n} \big|\sigma(t,X_t) - \sigma(t_n,{X}^\pi_{{n-1}})\big|^2 dt \right]
%   + O(h^2)  \\
% &\le \mathbb{E}\big|\delta X_{{n-1}}^\pi\big|^2 
%    + \int_{t_{n-1}}^{n_i} 
%      \mathbb{E}\left[ \left|\sigma(t,X_t) - \sigma(t_{n-1},X_{n-1})\right|^2 \right] dt \\
% &\quad + \int_{t_{n-1}}^{t_n} 
%      \mathbb{E}\!\left[ \left| \sigma(t_{n-1},X_{n-1}) - \sigma(t_{n-1},{X}_{{n-1}}^\pi) \right|^2 \right] dt 
%    + O(h^2) \\
% &\le \left(1 + \sigma_x h \right)\mathbb{E}\big|\delta X_{{n-1}}^\pi\big|^2 +  O(h^2) \\
% &\le e ^ {\sigma_x} \left(\mathbb{E}\big|\delta X_{{0}}^\pi\big|^2 +  O(h) \right) \\
% & = C h
% \end{aligned}
% \end{equation}
% \end{proof}

\begin{proposition}[Regularity results on $(X, Y^B, Z^B)$; Propositions 3.1 and 3.2 in \cite{chassagneux2016numerical}]
\label{prop:temporal_regularity}
Under the assumptions of Theorem \ref{th_conv_HJB}, the following regularity bounds hold:

\textbf{(Y-component)} For all $p \leq 1$:
\begin{equation}\label{eq:Y_estimate}
\sup_{0 \leq j \leq N-1} \mathbb{E}\left[ \sup_{t_j \leq s \leq t_{j+1}} |Y_s^B - Y_{t_j}^B|^{2p} \right] \leq C_p h^p.
\end{equation}

\textbf{(Z-component)} For all $p \geq 1$:
\begin{equation}\label{eq:Z_estimate}
\sum_{n=0}^{N-1}\mathbb{E}\left[\left(\int_{t_n}^{t_{n+1}} |Z_s^B - \bar{Z}_n^B|^2 ds \right)^p \right] \leq C_p h^p,
\end{equation}
where $\bar{Z}_n^B = \frac{1}{h}\int_{t_n}^{t_{n+1}}Z_s^B ds$.
\end{proposition}

\begin{lemma}\label{lemma:Z_n^B}
For any {$0 \le n \le N-1$}, we have
\[
\mathbb{E}_{n} \left|\tilde Z_n^\pi - \bar Z_n^B\right|
\;\le\; C h^{1/2}.
\]
\end{lemma}

\begin{proof}
By definition,
\begin{equation}\label{eq:def_tilde_Z}
\bar Z_n^B := \frac{1}{h}\, \mathbb{E}_{n}\!\left[\int_{t_n}^{t_{n+1}} Z_s^B ds \right], 
\qquad
\tilde Z_n^\pi := \mathbb{E}_{n}\!\left[\frac{1}{h} Y_{t_{n+1}}^B\,(W_{{n+1}}-W_{n}) \right].
\end{equation}
thanks to assumptions on $f^B$ ($f^B$ is B-Lipschitz-continuous with respect to $z$), and Cauchy–Schwarz inequality, for $n \le N$,
\begin{align*}
h \mathbb{E}_{n}\!\big[\,|\tilde Z_n^\pi - \bar Z_n^B|^2 \,\big]
&= h \mathbb{E}_{n}\!\left[ \Bigg|\mathbb{E}_{n}\!\left[ \int_{t_n}^{t_{n+1}} 
     f^B(s,X_s,Y_s^B,Z_s^B)\, ds \;\frac{W_{t_{n+1}}-W_{t_n}}{h_n} \right]\Bigg|^2 \right] \\
&\le h \mathbb{E}_{n}\!\left[ \int_{t_n}^{t_{n+1}} 
     \big| f^B(s,X_s,Y_s^B,Z_s^B)\big|^2 \, ds \right] \\
&\le C \Bigg( h^2 + (1+B^2 )h\, \mathbb{E}_{n}\!\left[ \int_{t_n}^{t_{n+1}} |Z_s^B|^2 ds \right]\Bigg).
\end{align*}
\end{proof}

\begin{proof}[Proof of Theorem \ref{th:DBSDE_conv_PBTZ_brief}]
Define $\delta {Y}_n^\pi = \widetilde Y_n^\pi - Y^\pi_n$, $\delta {Z}_n^\pi = \widetilde Z_n^\pi - Z^{\theta, \pi}_n$. By the system \eqref{eq:BTZ_system} and \eqref{system}, we have
\begin{equation}
\mathbb{E}\!\left[ \delta Y_n^\pi \right] 
= \mathbb{E}_n\!\big[ \delta Y_{n+1} ^\pi
  + h \big( f(t_n,X_n^\pi, Y_n^\pi, Z_n^{\theta, \pi}) - f^B(t_n,{X}_n^\pi, \widetilde{Y}_n^\pi, \widetilde{Z}_n^\pi) \big) \big] 
  + \mathbb{E}_n\!\left[ \Upsilon_n^Y \right] .
\end{equation}
and
\begin{equation}
\mathbb{E}\!\left| \delta Y_{n+1}^\pi  \right| 
\le \mathbb{E}_n\!\big| \delta Y_{n}^\pi \big| 
  + \mathbb{E}_n\! \big|   f(t_n,X_n^\pi, Y_n^\pi, Z_n^{\theta, \pi}) - f^B(t_n,{X}_n^\pi, \widetilde{Y}_n^\pi, \widetilde{Z}_n^\pi)  \big| h 
  + \mathbb{E}_n\!\left| \Upsilon_n^Y \right| .
\end{equation}
The key technical challenge lies in controlling the perturbation term $\mathbb{E}_n|\Upsilon_n^Y|$. We decompose:
\begin{align}
\mathbb{E}_n|\Upsilon_n^Y| &= \mathbb{E}_n|I_n^1| + \mathbb{E}_n|I_n^2| + \mathbb{E}_n|I_n^3|, \\
\mathbb{E}_n|I_n^1| &= \mathbb{E}_n\left|\int_{t_n}^{t_{n+1}} \big(f^B(s,X_s,Y_s^B,Z_s^B) - f^B(t_n,X_n^\pi,Y_s^B,Z_s^B)\big) ds\right|, \\
\mathbb{E}_n|I_n^2| &= \mathbb{E}_n\left|\int_{t_n}^{t_{n+1}} \big(f^B(t_n,X_n^\pi,Y_s^B,Z_s^B) - f^B(t_n,X_n^\pi,Y_{t_n}^B,Z_s^B)\big) ds\right|, \\
\mathbb{E}_n|I_n^3| &= \mathbb{E}_n\left|\int_{t_n}^{t_{n+1}} \big(f^B(t_n,X_n^\pi,Y_{t_n}^B,Z_s^B) - f^B(t_n,X_n^\pi,Y_{t_n}^B,\widetilde{Z}_n^\pi)\big) ds\right|.
\end{align}
Applying the Lipschitz properties of $f^B$ with respect to the spatial and temporal variables, combined with Lemma \ref{lemma:Truncation_X_Convergence} and Proposition \ref{prop:temporal_regularity}, we obtain
$$\mathbb{E}_n \!\left| I_1 \right| \le C_1 h^{\frac{3}{2}}, \quad \mathbb{E}_n \!\left| I_2 \right| \le C_2 h^{\frac{3}{2}},$$
for positive constants $C_1$ and $C_2$ independent of the discretization parameter $h$.
The proof of $I_3$ constitutes the most technically demanding component, requiring careful analysis of the quadratic structure and projection properties.
\begin{equation}
\begin{aligned}
\mathbb{E}_n \!\left| I_3 \right| & =\mathbb{E}_n \!\left| \int_{t_n}^{t_{n+1}} 
      \Big( f^B\big(t_n, X_n^\pi, Y_n^B, Z_s^B\big) 
           - f^B\big(t_n, {X}_n^\pi, Y_n^B, \widetilde{Z}_n^\pi\big) \Big)\, ds \right|\\
&  \le \mathbb{E}_n \!\left| \int_{t_n}^{t_{n+1}} 
      \Big( f^B\big(t_n, X_n^\pi, Y_n^B, Z_s^B\big) 
           - f^B\big(t_n, {X}_n^\pi, Y_n^B, \bar{Z}_n^B\big) \Big)\, ds \right|\\
  & \quad + \mathbb{E}_n \!\left| \int_{t_n}^{t_{n+1}} 
      \Big( f^B\big(t_n, X_n^\pi, Y_n^B,  \bar Z_n^B\big) 
           - f^B\big(t_n, {X}_n^\pi, Y_n^B, \widetilde{Z}_n^\pi\big) \Big)\, ds \right|\\
&  \le C_f \left[\int_{t_n}^{t_{n+1}} \mathbb{E}_n\! \left| Z_s^B - \bar{Z}_n^B \right|^2 ds + \int_{t_n}^{t_{n+1}} \mathbb{E}_n\! \left| \bar Z_n^B - \widetilde{Z}_n^\pi \right|^2 ds \right].
\end{aligned}
\end{equation}
Exploiting the truncated Lipschitz properties of $f^B$ and applying the regularity estimates from Proposition \ref{prop:temporal_regularity} and Lemma \ref{lemma:Z_n^B}, we derive
$$\mathbb{E}_n \!\left| I_3 \right| \le C_3 h^{\frac{3}{2}},$$
where the constant $C_3$ is independent of $h$.
consequently, the error evolution satisfies
% \begin{equation}
% \begin{aligned}
% \mathbb{E}\!\left| \delta Y_{n+1}^\pi  \right| 
% &\le \mathbb{E}_n\!\big| \delta Y_{n}^\pi \big| 
%   + \mathbb{E}_n\! \big|   f(t_n,X_n^\pi, Y_n^\pi, Z_n^{\theta, \pi}) - f(t_n,{X}_n^\pi, \widetilde{Y}_n^\pi, \widetilde{Z}_n^\pi)  \big| h 
%   + \mathbb{E}_n\!\left| \Upsilon_n^Y \right| \\
% &\le C_f \mathbb{E}_n\!\big| \delta Y_{n}^\pi \big|  + C_f \mathbb{E}_n\!\big| \left(Z^{\theta,\pi}_{n} - \widetilde{Z}_n^\pi\right)^2  \big|h + O(h^\frac{3}{2})\\
% &\le C_f (\mathbb{E}_0\!\big| Y_0 - \theta_{u_0} \big| + \sum_{n=0}^{n}\mathbb{E}_n\!\big| \left(Z^{\theta,\pi}_{n} - \widetilde{Z}_n^\pi\right)^2  \big|)h + O(h^{\frac{1}{2}}).
% \end{aligned}
% \end{equation}
{
\begin{equation}\label{eq:recursion-corrected_1}
\begin{aligned}
\mathbb{E}\!\left| \delta Y_{n+1}^\pi \right| 
&\le \mathbb{E}_n\!\left| \delta Y_{n}^\pi \right|
+ \mathbb{E}_n\!\Big| f(t_n,X_n^\pi, Y_n^\pi, Z_n^{\theta, \pi})
      - f^B(t_n,X_n^\pi, \widetilde{Y}_n^\pi, \widetilde{Z}_n^\pi)\Big|\, h
+ \mathbb{E}_n\!\left| \Upsilon_n^Y \right| \\[2mm]
&\le (1+ C_f h)\,\mathbb{E}_n\!\left| \delta Y_{n}^\pi \right|
+ C_f\,\mathbb{E}_n\!\left[ \left| Z^{\theta,\pi}_{n} - \widetilde{Z}_n^\pi \right|^2 \right] h
+ O(h^{3/2}) 
\end{aligned}
\end{equation}
By applying the discrete Gronwall inequality, for $n\le N-1$, we obtain:
\begin{equation}\label{eq:recursion-corrected}
\begin{aligned}
\mathbb{E}\!\left| \delta Y_{n+1}^\pi \right| \le e^{C_f T}\!\left(\mathbb{E}_0\!\left| Y_0 - \theta_{u_0} \right|
+ \sum_{k=0}^{n}\mathbb{E}_k\!\left[ \left| Z^{\theta,\pi}_{k} - \widetilde{Z}_k^\pi \right|^2 \right] h\right)
+ O(h^{1/2}) .
\end{aligned}
\end{equation}
}
\end{proof}

\section*{{Appendix D: A posteriori estimation of the simulation error for HJB-type equation}}

\begin{lemma}
\label{lemma:martingale}
Let $0 \leq s_1 < s_2$, given $Q \in L^2(\Omega, \mathcal{F}_{s_2}, \mathbb{P})$, by the martingale representation theorem, there exists an $\mathcal{F}_t$-adapted process $\{H_s\}_{s_1 \leq s \leq s_2}$ such that $\int_{s_1}^{s_2} E|H_s|^2 \, ds < \infty$ and $Q = E[Q|\mathcal{F}_{s_1}] + \int_{s_1}^{s_2} H_s \, dW_s$. Then we have $E[Q(W_{s_2} - W_{s_1})|\mathcal{F}_{s_1}] = E[\int_{s_1}^{s_2} H_s \, ds|\mathcal{F}_{s_1}]$.
\end{lemma}
The lemma is from \cite{Convergence-of-the-deep-BSDE-method}. 

Rewite \eqref{eq:perturbed_BTZ}, we have
\begin{equation}\label{app_eq:pf_tilde_Y}
\widetilde{Y}_{n+1}^\pi = \widetilde{Y}_n^\pi - \int_{t_n}^{t{n+1}} f^B(s,X_s,Y_s^B,Z_s^B)\,ds + \int_{t_n}^{t{n+1}} Z_s^B \, dW_s
\end{equation}
we apply Lemma \ref{lemma:martingale} to \eqref{eq:def_tilde_Z}, \eqref{app_eq:pf_tilde_Y} and get
\[
\tilde Z_n^\pi = \frac{1}{h} \mathbb E \left[ \int_{t_n}^{t_{n+1}} Z_t^B dt  |\mathcal{F}_{n}\right],
\]
which implies, by the Cauchy inequality,
\begin{equation}\label{app_eq_pf:Z-Z}
\begin{aligned}
\mathbb E|\widetilde Z_n^\pi-Z_{n}^{\theta,\pi}|^2 h &=  \sum_{k=1}^d \mathbb E|(\widetilde Z_n^\pi-Z_{n}^{\theta,\pi})_k|^2 h 
=  \sum_{k=1}^d \frac{1}{h} \mathbb E\left|\mathbb E\left[\int_{t_n}^{t_{n+1}} (Z_t^B - Z_{n}^{\theta,\pi})_k \, dt \Big| \mathcal{F}_{t_n}\right]\right|^2 \\
&\leq  \sum_{k=1}^d \frac{1}{h} E\left|\int_{t_n}^{t_{n+1}} (Z_t^B - Z_{n}^{\theta,\pi})_k \, dt\right|^2 \leq  \sum_{k=1}^d \int_{t_n}^{t_{n+1}} E|(Z_t^B - Z_{n}^{\theta,\pi})_k|^2 \, dt \\
&= \int_{t_n}^{t_{n+1}} E|(Z_t^B - Z_{n}^{\theta,\pi})|^2 \, dt ,
\end{aligned}
\end{equation}

\begin{proof} [Proof of Theorem \ref{Th:estimation_HJB}]
Define $\delta {Y}_n^\pi = \widetilde Y_n^\pi - Y^\pi_n$,
\begin{equation}\label{app_eq:deltaY}
\delta Y_{n+1}^\pi=
\delta Y_{n}^\pi
-\Big(
f^B(t_n, X_n^\pi, \widetilde Y_n^\pi, \widetilde Z_n^\pi)-
f(t_n, X_{t_n}^\pi, Y_{t_n}^\pi, Z_{t_n}^{\theta,\pi})
\Big)h
+\Big(\int_{t_n}^{t_{n+1}} Z_s^B\, -Z_{t_n}^{\theta,\pi} dW_s\Big)
+\Upsilon_n^Y ,
\end{equation}
where $\Upsilon_n^Y$ is from equation \eqref{eq:Upsilon}.
From equation \eqref{app_eq:deltaY}, by \textbf{$H_1$}, \textbf{$H_3$} in Assumption \ref{assumption_1}, Assumption \ref{assumption_3} Assumption \ref{ass:quadratic}, and the root-mean square and geometric
mean inequality (RMS-GM inequality),
for any $\lambda_1 >0,$ we have
\begin{equation}\label{app_eq:delta_Y2_1}
\begin{aligned}
\mathbb E|\delta Y_{n+1}^\pi|^2
&= \mathbb E|\delta Y_{n}^\pi|^2 + \mathbb E \big[|
f^B(t_n, X_n^\pi, \widetilde Y_n^\pi, \widetilde Z_n^\pi)-
f^B(t_n, X_{n}^\pi, Y_{n}^\pi, \widetilde Z_n^\pi) |^2\big]h^2 +\mathbb E|\Upsilon_n^Y|^2 +\int_{t_n}^{t_{n+1}} \mathbb E|Z_s^B-Z_{n}^{\theta,\pi}|^2\,ds    \\
&\qquad -2\,\mathbb E\big[(
f^B(t_n, X_n^\pi, \widetilde Y_n^\pi, \widetilde Z_n^\pi)-
f^B(t_n, X_{n}^\pi, Y_{n}^\pi, \widetilde Z_n^\pi) \delta Y_n^\pi \big]h -2 \mathbb E|\Upsilon_n^Y \delta Y_{n}^\pi|\\
&\qquad -2\,\mathbb E\big[(
f^B(t_n, X_n^\pi, \widetilde Y_n^\pi, \widetilde Z_n^\pi)-
f^B(t_n, X_{n}^\pi, Y_{n}^\pi, \widetilde Z_n^\pi) \Upsilon_n^Y \big]h\\
&\ge \ \mathbb E|\delta Y_{n}^\pi|^2
+\int_{t_n}^{t_{n+1}} \mathbb E |Z_s^B-Z_{t_n}^{\theta,\pi}|^2\,ds  -2\mathbb E\Big[\left(
f^B(t_n, X_n^\pi, \widetilde Y_n^\pi, \widetilde Z_n^\pi)-
f^B(t_n, X_{n}^\pi, Y_{n}^\pi, \widetilde Z_n^\pi)
\right) \delta Y_{n}^\pi \Big]h\\
&\qquad -2\mathbb E\Big[\Big(
f^B(t_n, X_n^\pi, Y_n^\pi, \widetilde Z_n^\pi)-
f(t_n, X_{n}^\pi, Y_{n}^\pi, Z_{n}^{\theta,\pi})
\Big) \delta Y_{n}^\pi \Big]h +O(h^{3/2})\\
&\ge \ \mathbb E|\delta Y_{n}^\pi|^2
+\int_{t_n}^{t_{n+1}} \mathbb E |Z_s^B-Z_{t_n}^{\theta,\pi}|^2\,ds  -2 C_f \mathbb E|\delta Y_{n}^\pi|^2 h\\
&\qquad -\Big[ \lambda_1 \mathbb E|\delta Y_{n}^\pi|^2 + \frac{1}{\lambda_1} \left( C_f (1 + |\varphi^B(\widetilde Z_n^\pi)|^2 + |Z_{n}^{\theta,\pi}|^2 ) |\widetilde Z_n^\pi - Z_{n}^{\theta,\pi}|^2 \right) \Big]h  + O(h^{3/2}).
\end{aligned}
\end{equation}
Plugging it into \eqref{app_eq_pf:Z-Z} gives us
\begin{equation}\label{app_pf_eq:1}
\begin{aligned}
\mathbb E|\delta Y_{n+1}^\pi|^2
&\ge \left[1-(2C_f + \lambda_1)h \right] \mathbb E|\delta Y_{n}^\pi|^2 + \left[1-\frac{C_f}{\lambda_1}(1+ 2B^2)\right] h E|\delta Z_{n}^\pi|^2 + O(h^{3/2}).
\end{aligned}
\end{equation}
Then there exists a constant $C$, for any $\lambda_1 \ge C_f(1+2B^2)$ and sufficiently small $h$ satisfying $(2C_f + \lambda_1)h < 1$, we have
\[
\begin{aligned}
\mathbb E|\delta Y_{n}^\pi|^2 &\le e^{-h^{-1}ln\left[1-(2C_f + \lambda_1)h\right](N-n)h}\left[\mathbb E|\widetilde Y_N^\pi-Y_N^\pi|^2 + O(h^{1/2}) \right]\\
&\le C \left[\mathbb E|\widetilde Y_N^\pi-Y_N^\pi|^2 + O(h^{1/2})\right].
\end{aligned}
\]

\end{proof}

\begin{proof}[Proof of Theorem \ref{th_conv_HJB}]

Combine Lemma \ref{lemma:Truncation_Convergence_brief} and Theorem \ref{Th:estimation_HJB}, there exists a constant $C_1' > 0$ such that
\[
\begin{aligned}
\mathbb E|Y_{t_n} -Y_{n}^\pi|^2 &\le e^{-h^{-1}ln\left[1-(2C_f + \lambda_1)h\right](N-n)h}\left[\mathbb E|g(X_T^\pi)-Y_N^\pi|^2 + O(h^{1/2}) +   \mathbb E|g(X_T^\pi)- Y_T^B|^2 \right] + \mathbb E|Y_{t_n} -Y_{t_n}^B|^2\\
&=e^{-h^{-1}ln\left[1-(2C_f + \lambda_1)h\right](N-n)h}\left[\mathbb E|g(X_T^\pi)-Y_N^\pi|^2 + O(h^{1/2}) +C_p D_\beta B^{-\tfrac{\beta}{2\bar{q}}} \right] + C_p D_\beta B^{-\tfrac{\beta}{2\bar{q}}}\\
&\le C_1' (h^{1/2}+\mathbb E|g(X_T^\pi)-Y_N^\pi|^2 + C_p D_\beta B^{-\tfrac{\beta}{2\bar{q}}}).
\end{aligned}
\]
Specially, 
\[
\begin{aligned}
\mathbb E|u(0,\xi) - \theta_{u_0}|^2 \le C_1' (h^{1/2}+\mathbb E|g(X_T^\pi)-Y_N^\pi|^2 + C_p D_\beta B^{-\tfrac{\beta}{2\bar{q}}})
\end{aligned}.
\]
Combine Lemma \ref{lemma:Truncation_Convergence_brief} and Theorem \ref{th:DBSDE_conv_PBTZ_brief}, there exists a constant $C_2' > 0$ such that
\begin{equation}
\begin{aligned}
\sup_{0 \le n \le N} \, \mathbb{E}\left|Y_{t_n} - Y^{\pi}_n\right|
\;\le\; &\sup_{0 \le n \le N} \, \mathbb{E}\left|\delta Y_n^\pi\right| + \sup_{0 \le n \le N} \, \mathbb{E}\left|Y_{t_n}^B - Y_{t_n}\right|     \\
\;\le\; &C_2'\Bigg( h^{\frac{1}{2}} +
    \mathbb{E}\left|Y_{0} - \theta_{u_0}\right|^{2}
    + \sum_{n=0}^{N-1} \mathbb{E}\left|Z^{\theta,\pi}_{n} - \tilde{Z}^{\pi}_n\right|^{2} h
\Bigg)
\;+\; \sqrt {C_p D_{\beta}} B^{-\frac{\beta}{4q}}.
\end{aligned}
\end{equation}
Specially, 
\begin{equation}
\begin{aligned}
\mathbb{E}\left|{g(X_T^\pi)-Y^\pi_N}\right| \leq C_2'\left(h^{\frac{1}{2}} +\mathbb{E}\left|Y_{0} - \theta_{u_0}\right|^{2} + \sum_{n=0}^{N-1} \mathbb{E}\left|\widetilde{Z}_n^\pi-Z^{\theta,\pi}_{n}\right|^2 h \right)
+ \sqrt {C_p D_{\beta}} B^{-\frac{\beta}{4q}}.
\end{aligned}
\end{equation}
After the neural network $\phi_n^\theta\left(t_n,X_{t_n}^\pi\right)$ has been sufficiently trained, we obtain
\begin{equation}
\begin{aligned}
\inf_{\theta_{{u_0}},\theta_{\nabla{u_0}} \in \mathcal{N}_0, \phi_n \in \mathcal{N}_n} \mathbb{E}\left|{g(X_T^\pi)-Y^\pi_N}\right|  &\leq C_2'\left(h^{\frac{1}{2}} + \inf _{\theta_{{u_0}},\theta_{\nabla{u_0}} \in \mathcal{N}_0} \mathbb{E}\left|Y_{0} - \theta_{u_0}\right|^{2} +  \mathbb{E}\left| \widetilde Z_0^\pi-Z_0^{\theta,\pi} \right|^2 h \right.\\
& \left. \qquad  +  \inf _{\phi_n^\theta \in \mathcal{N}_n} \sum_{n=1}^{N-1} \mathbb{E}\left|\widetilde{Z}_n^\pi-Z_n^{\theta,\pi} \right|^2 h \right)
+ \sqrt {C_p D_{\beta}} B^{-\frac{\beta}{4q}}.
\end{aligned}
\end{equation}

\end{proof}

\end{document}